\documentclass[aps,prd,superscriptaddress,floatfix,nofootinbib,notitlepage,12pt]{revtex4-1}

\usepackage{aas_macros}
\usepackage{graphicx}
\usepackage{multirow}
\usepackage{color}
\usepackage{slashed}
\usepackage{comment}
\usepackage{ulem}
\usepackage{soul}
\setstcolor{red}
\usepackage{amsmath,amssymb,amsfonts}
\usepackage{verbatim}
\usepackage{bm}
\usepackage{color}
\usepackage{ulem}
\usepackage{mathtools}
\usepackage{colortbl}
\usepackage[dvipsnames]{xcolor}
\usepackage{slashed}
\usepackage{braket}
\usepackage{adjustbox}
\usepackage{cancel}
\usepackage{makecell}

\usepackage{ulem}
\usepackage{soul}
\setstcolor{red}

\usepackage[colorlinks]{hyperref}
\hypersetup{
	colorlinks=true,
	linkcolor=blue,
	filecolor=blue,      
	urlcolor=blue,
	citecolor=magenta,
	pdfpagemode=FullScreen
} 

\newcommand{\orcid}[1]{\hspace{1mm}\href{https://orcid.org/#1}{\includegraphics[height=0.3cm,keepaspectratio]{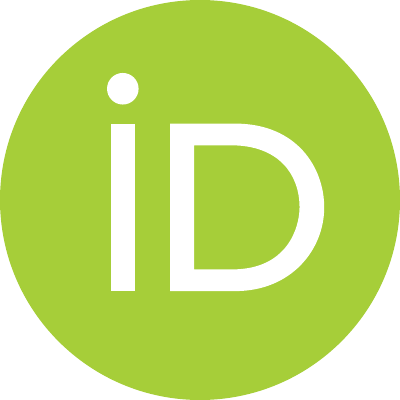}}}

\begin{document}
	
	\title{Searching for charged Higgs bosons via $e^+ e^- \to H^\pm W^\mp S$ at the ILC}
	\author{Brahim Ait Ouazghour\orcid{0009-0006-1419-969X}}
	\email{b.ouazghour@gmail.com}
	\affiliation{LPHEA, Physics Department, FSSM, Cadi Ayyad University, P.O.B. 2390 Marrakech, Morocco}
	\author{Abdesslam Arhrib\orcid{0000-0001-5619-7189}}
	\email{aarhrib@gmail.com}
	\affiliation{Abdelmalek Essaadi University, FST Tanger B.P. 416, Morocco}
	\affiliation{Department of Physics and CTC, National Tsing Hua University, Hsinchu, Taiwan 300}
	\author{Kingman Cheung\orcid{0000-0003-2176-4053}}
	\email{cheung@phys.nthu.edu.tw}
	\affiliation{Department of Physics and CTC, National Tsing Hua University, Hsinchu, Taiwan 300}
	\affiliation{Division of Quantum Phases and Devices, School of Physics,
		Konkuk University, Seoul 143-701, Republic of Korea}

	\author{Es-said Ghourmin\orcid{0009-0007-1597-8537}}
	\email{s.ghourmin123@gmail.com}
	\affiliation{Laboratory of Theoretical and High Energy Physics (LPTHE), Faculty of Science, Ibnou Zohr University, B.P 8106, Agadir, Morocco}
	
	\author{Mohamed Krab\orcid{0000-0003-2366-0493}}
	\email{mkrab@hep1.phys.ntu.edu.tw}
	\affiliation{Department of Physics, National Taiwan University, Taipei 10617, Taiwan}

	\author{Larbi Rahili\orcid{0000-0002-1164-1095}}
	\email{rahililarbi@gmail.com}
	\affiliation{Laboratory of Theoretical and High Energy Physics (LPTHE), Faculty of Science, Ibnou Zohr University, B.P 8106, Agadir, Morocco}	
	\date{\today}
	\begin{abstract}
		We investigate the phenomenology of the charged Higgs boson at the International Linear Collider (ILC) within the framework of the type-X Two-Higgs Doublet Model (2HDM), where a light charged Higgs boson, with a mass around 200 GeV or even smaller than top quark mass, is still being consistent with flavor physics data as well as with the colliders experimental data. 
		In the theoretically and experimentally allowed parameter space, the $e^+ e^- \to H^\pm W^\mp S$ (with $S = H, A$) production processes can yield signatures with event rates larger than those from 
		$e^+ e^- \to H^+ H^-$ and offer sensitivity to the Higgs mixing parameter $\sin(\beta-\alpha)$. 
		We consider the bosonic $H^\pm \to W^\pm S$ decays, where the neutral scalar $S$ further decays into a pair of tau leptons. 
		We show, through a detector-level Monte Carlo analysis, that the resulting $[\tau\tau][\tau\tau] WW$ final state could be seen at the ILC with at least 500 GeV center-of-mass energy and 500 fb$^{-1}$ of luminosity.
		
	\end{abstract}
	
	\maketitle
	
	\newpage
	\section{Introduction}

	The discovery of the Standard Model (SM) Higgs boson by ATLAS \cite{ATLAS:2012yve} and CMS \cite{CMS:2012qbp} experiments has completed the particle spectrum defined by elementary particle physics. However, many questions remain unanswered. These include theoretical challenges, such as the mechanism that stabilizes the electroweak scale and the generation of neutrino masses, as well as empirical issues like the nature of dark matter and the matter-antimatter asymmetry. As a result, there is a strong motivation to investigate theories beyond the SM (BSM) that operate near the TeV scale \cite{Ait-Ouazghour:2020slc,Grzadkowski:2011jks,karahan2014effects,darvishi2018implication,Ouazghour:2018mld}. Many BSM theories naturally feature an extended Higgs sector. A prominent example of this is the two Higgs doublet model (2HDM), which adds an additional electroweak Higgs doublet, enriching the theory with a variety of new Higgs boson phenomena and implications for flavor physics. Intense searches for these novel Higgs bosons have been conducted at colliders, particularly at the Large Hadron Collider (LHC), which has played a significant role in these efforts (for a comprehensive overview, see Ref. \cite{Kling:2020hmi} and the references therein). The lack of observed signals has resulted in current limits on the mass and couplings of these non-SM Higgs bosons. The High Luminosity LHC (HL-LHC) is expected to enhance several earlier measurements and may provide evidence for new physics. However, to advance the precise Higgs measurement program initiated at the LHC, it is essential to establish a controlled environment, such as an electron-positron Higgs factory. This would facilitate a thorough investigation of the properties of the recently discovered Higgs boson, similar to those in the SM, and could even enable the discovery of new particles. Several projects involving 
	$e^+e^-$ machines are currently in the planning stages, including initiatives like the Circular Electron Positron Collider (CEPC) \cite{An:2018dwb}, the Compact Linear Collider (CLIC) \cite{CLICPhysicsWorkingGroup:2004qvu,Aicheler:2012bya}, the Future Circular Collider (FCC-ee) \cite{FCC:2018evy,TLEPDesignStudyWorkingGroup:2013myl}, and the International Linear Collider (ILC) \cite{LCCPhysicsWorkingGroup:2019fvj,Moortgat-Pick:2015lbx}.

	These projects aim to provide an ideal setting for intricate investigations into the properties of the Higgs boson, and potentially discover new particles. Recent investigations have indeed highlighted the remarkable prospects these colliders present in exploring the electroweak sector. This includes endeavors such as precise measurements of Higgs boson couplings \cite{Han:2020pif}, the detection of electroweak dark matter \cite{Han:2020uak,Belfkir:2023vpo}, study of neutrinos \cite{Jueid:2023qcf,Jana:2023ogd} and the potential discovery of other BSM particles \cite{Costantini:2020stv,Bandyopadhyay:2024plc,Han:2021udl}.
	
	Existence of charged Higgs bosons is a generic prediction of multiple extensions of the SM Higgs sector, such as the 2HDM. The charged Higgs bosons can be produced and explored at the LHC. We refer to Ref. \cite{Akeroyd:2016ymd} for an extensive review on charged Higgs phenomenology at the LHC. A light $H^\pm$ can be copiously produced from $t\bar{t}$ production followed by $t\to b H^+$ if kinematically allowed (i.e. $m_{H^\pm} < m_t - m_b$).
	When its exceeds the top quark mass, the processes $gb\to t H^-$ and  $gg \to  \bar{t} b H^+$ \cite{Barger:1993th} are typically used for $H^\pm$ searches.
	
	Charged Higgs production  can proceed at ILC 
	\cite{ILC:2013jhg,ECFADESYLCPhysicsWorkingGroup:2001igx} or 
	CLIC \cite{CLICPhysicsWorkingGroup:2004qvu} through $e^+e^-\to \gamma^*, Z^* \to H^+H^-$ or $e^+ e^- \to \tau^{\pm} \nu_{\tau }H^{\mp}$ and $e^+ e^- \to t \bar{b} H^{-}$ \cite{Ouazghour:2024twx,Komamiya:1988rs,Kanemura:2000cw,Brignole:1991fw} and the loop mediated process $e^+e^- \to W^\pm H^\mp$ \cite{Arhrib:1999rg,Kanemura:1999tg}. At high energy, one can also have vector boson fusion:   $e^+ e^- \to \nu \bar{\nu}H^{+}H^-$\cite{Ouazghour:2023plc}. The production of a charged Higgs boson at muon colliders is rather similar to $e^+e^-$, it can proceed through several processes. Production processes $\mu^+ \mu^- \to H^+H^-$, $\mu^+ \mu^- \to H^{\pm}W^{\mp}$, and $\mu^+ \mu^- \to \nu \bar{\nu} H^+ H^-$ have been studied in Ref. \cite{Ouazghour:2023plc}. 
	
	Charged Higgs bosons have been searched for at LEP \cite{DELPHI:2003eid} through $e^+ e^- \to H^+ H^-$, followed by the fermionic $H^\pm \to \tau\nu$ and $H^\pm \to cs$ decays, where a limit of $m_{H^\pm} \geq  80$ GeV was obtained assuming $H^\pm \to W^\pm A$ is absent.
	If the latter is present, a limit of $m_{H^\pm} \geq 72.5$~GeV (2HDM type-I) was obtained \cite{ALEPH:2013htx} assuming $m_A = 12$ GeV. 
	At the Tevatron, charged Higgs bosons have also been searched for in top quark decays with subsequent $\tau\nu$ or $cs$ decays \cite{CDF:2005acr,D0:2009hbc}. These searches set upper limits on BR$(t \to bH^\pm)$ for different $H^\pm$ decay scenarios. 
	At the LHC, charged Higgs bosons have been searched for in $H^\pm \to \tau\nu$~\cite{ATLAS:2012nhc,ATLAS:2012tny,ATLAS:2016avi, ATLAS:2014otc, ATLAS:2018gfm,ATLAS:2024hya,CMS:2012fgz,CMS:2014cdp,CMS:2014pea,CMS:2015lsf,CMS:2019bfg,CMS:2018ect}, $cs$ \cite{ATLAS:2013uxj,CMS:2015yvc,CMS:2020osd,ATLAS:2024oqu}, $cb$\footnote{ATLAS reported a charged Higgs excess at $m_{H^\pm} = 130$ GeV \cite{ATLAS:2023bzb}; see, e.g., Ref. \cite{Arhrib:2024sfg} for a possible interpretation within 2HDM type-III.}~\cite{CMS:2018dzl,ATLAS:2023bzb}, $tb$ \cite{CMS:2014pea,ATLAS:2015nkq,ATLAS:2018ntn,ATLAS:2021upq,CMS:2019rlz,CMS:2020imj} decays. 
	All the aforementioned searches are conventional without considering bosonic decays.
	However, if bosonic decay channels $H^\pm \to W^{\pm (*)} S$ ($S = h, H, A$) are kinematically accessible, they can offer distinctive signatures and could be promising avenues for discovery \cite{Arhrib:2016wpw,Arhrib:2020tqk,Bahl:2021str,Arhrib:2021xmc,Arhrib:2021yqf,Mondal:2021bxa,Arhrib:2022inj,Krab:2022lih}.
	A few experimental efforts have begun probing these channels \cite{CMS:2019idx,ATLAS:2021xhq,CMS:2022jqc,ATLAS:2024rcu}, but they remain limited.
	See, e.g., Ref. \cite{Li:2024kpd} for a recent detailed study of collider search limits on charged Higgs bosons in the four types of the 2HDM.
	
	In this study, we investigate the production of charged Higgs bosons in association with a $W$ boson and a non-SM Higgs boson, i.e. $e^+ e^- \to H^\pm W^{\mp} S$, where $S=H$ or $A$, within the framework of the type-X 2HDM. 
	Our numerical results are provided after thoroughly exploring the 2HDM parameter space, adhering to various theoretical (perturbative unitarity, perturbativity, and vacuum stability) and experimental (derived from SM-like Higgs boson discovery data, BSM Higgs bosons exclusion data, electroweak precision tests (EWPT), and flavor physics) constraints. We perform a comprehensive Monte Carlo analysis and gauge the sensitivity at center-of-mass energies of 500, 1000 and 1500 GeV of the ILC. 
		
	\section{The 2HDM}
	\label{sec:model}
	\label{subsec:review}
	In the 2HDM, in addition to the SM scalar doublet $\Phi_1$, an additionnal doublet $\Phi_2$ with hypercharge $+1$ is added to the Higgs sector, assuming that Charge-Parity (CP) is not spontaneously broken. The two Higgs scalar doublets can be parametrized as follows:
	\begin{equation}
	\Phi_1 = \left(
	\begin{array}{c}
	\phi_1^+ \\
	\phi_1^0 \\
	\end{array}
	\right)
	\quad {\rm and}\quad 
	\Phi_2 = \left(
	\begin{array}{c}
	\phi_2^+ \\
	\phi_2^0 \\
	\end{array}
	\right),
	\end{equation}
	with $\phi_1^0 = (v_1+\psi_1+ i \eta_1)/\sqrt{2}$, $\phi_2^0 = (v_2+\psi_2+ i \eta_2)/\sqrt{2}$, and $\sqrt{v_1^2+v_2^2}=v=246$ GeV. The general scalar potential invariant under the electroweak gauge group $SU(2)_L\times U(1)_Y$ can be expressed as \cite{Branco:2011iw}:
	\begin{eqnarray}
	V(\Phi_1,\Phi_2) &=& m_{11}^2 \Phi_1^\dagger\Phi_1+m_{22}^2\Phi_2^\dagger\Phi_2-[m_{12}^2\Phi_1^\dagger\Phi_2+{\rm h.c.}] + \frac{\lambda_1}{2}(\Phi_1^\dagger\Phi_1)^2 + \frac{\lambda_2}{2}(\Phi_2^\dagger\Phi_2)^2\nonumber\\
	&+&\lambda_3(\Phi_1^\dagger\Phi_1)(\Phi_2^\dagger\Phi_2)
	+\lambda_4(\Phi_1^\dagger\Phi_2)(\Phi_2^\dagger\Phi_1) 
	+\left\{\frac{\lambda_5}{2}(\Phi_1^\dagger\Phi_2)^2+\rm{h.c}\right\},\label{pot1}
	\label{scalar_pot}
	\end{eqnarray}
	where $m_{11}^2$, $m_{22}^2$, and $m_{12}^2$ parameters as well as the $\lambda_{i}\,(i=1,2,3,4,5)$ couplings are assumed to be real to ensure that the potential is CP-conserving. A discrete $Z_2$ symmetry is also assumed in order to avoid Flavor Changing Neutral Currents (FCNCs) at the tree level. Such a $Z_2$ symmetry is softly broken by the bilinear term proportional to $m_{12}^2$ parameter. 
	
	After electroweak symmetry breaking, the 8 degrees of freedom initially present in the two Higgs doublet fields are reduced. Three of these degrees of freedom are absorbed by the Goldstone bosons, giving mass to the gauge bosons $W^\pm$ and $Z$. We are then left with five physical Higgs states: a pair of charged Higgs $H^\pm$, a CP-odd state $A$ and two CP-even states: $H$ and $h$ with $m_h < m_H$. One of the neutral CP-even Higgs can be identified as the 125 GeV Higgs-like particle observed at the LHC. 
	The combination $v^2=v_1^2+v_2^2=(2\sqrt{2} G_F)^{-1}$ can be used to fix one of the vacuum expectation values (vev) as a function of the Fermi constant $G_F$ and $\tan\beta$. Together with the two minimization conditions, the model has seven independent parameters:
	\begin{equation}
	\label{eq:modelpara}
	\alpha,\quad \tan\beta(= v_2/v_1),\quad  m_{h},\quad m_{H},\quad m_A,\quad 
	m_{H^\pm},\quad \mathrm{and} \quad m_{12}^2,
	\end{equation}
	where $\alpha$ and $\beta$ are respectively the CP-even and CP-odd mixing angles. 
	Here, we assume that $h$ is the observed SM-like Higgs boson at the LHC with $m_h=125.09$ GeV, therefore we are left with only six free parameters. 
	
	In the Yukawa sector, if we assume that both Higgs doublets couple to all fermions, like in the SM, we will end up with a large tree level FCNCs mediated by the neutral Higgs scalars. To prevent such large FCNCs at the tree level, the 2HDM needs to satisfy Paschos-Glashow-Weinberg theorem \cite{Paschos:1976ay,Glashow:1976nt} which asserts that all fermions with the same quantum numbers couple to the same Higgs multiplet. One can have four different types of Yukawa textures depending on how the doublets $\Phi_1$ and $\Phi_2$ interact with the fermions.
	In the 2HDM type-I, only the second doublet $\Phi_2$ interacts with all the fermions.
	In type-II, $\Phi_2$ interacts with up-type quarks and $\Phi_1$ interacts with the charged leptons and down-type quarks, while in type-Y (or flipped) down-type quarks couple to $\Phi_1$, while charged leptons and up-type quarks couple to $\Phi_2$.
	In type-X, $\Phi_2$ couples to quarks and $\Phi_1$ couples to leptons.
	
	In terms of the mass eigenstates of the neutral and charged Higgs bosons, the Yukawa interactions can be written as:
	\begin{eqnarray}
	-{\cal L}_Y &=& \sum_{f=u,d,\ell} \frac{m_f }{v} \left[ \kappa^f_h  \bar f f  h + \kappa^f_H \bar f f H - i \kappa^f_A \bar f \gamma_5 f A \right]   \label{eq:Yukawa_CH} \nonumber \\
	&&\qquad~ +\,\frac{\sqrt{2}}{v} \left[ \bar u_{i} V_{ij}\left( m_{u_i}  \kappa^{u}_A P_L + \kappa^{d}_A  m_{d_j} P_R \right)d_{j}  H^+ \right] \nonumber \\ 
	&& \qquad~ +\, \frac{\sqrt{2}}{v}  \bar \nu_L  \kappa^\ell_A m_\ell \ell_R H^+ +\rm{h.c.}, 
	\end{eqnarray} 
	where $\kappa^f_S$ ($S = h, H, A$) are the Yukawa couplings listed in Table \ref{coupIII} for the type-X 2HDM, and $V_{ij}$ are CKM matrix elements. In what follows, we use the notation of $s_x = \sin x$ and $c_x = \cos x$.
	
	\begin{table}
		\begin{center}
			\begin{tabular}{|c|c|c|c|c|c|c|c|c|c|}
				\hline 
				& $\kappa_h^u$ & $\kappa_h^d$ & $\kappa_h^l$ & $\kappa_H^u$ & $\kappa_H^d$ & $\kappa_H^l$ & $\kappa_A^u$ & $\kappa_A^d$ & $\kappa_A^l$ \\
				\hline 
				
				type-X & $c_\alpha/s_\beta$ & $c_\alpha/s_\beta$& $-s_\alpha/c_\beta$ & $s_\alpha/s_\beta$ & $s_\alpha/s_\beta$ & $c_\alpha/c_\beta$ & $c_\beta/s_\beta$ & 
				$-c_\beta/s_\beta$ & $\tan\beta$ \\ \hline
			\end{tabular}
		\end{center}
		\caption{Yukawa couplings of the $h$, $H$, and $A$ Higgs bosons to fermions in 2HDM type-X.} 
		\label{coupIII}
	\end{table} 
	
	The reduced coupling of the lighter Higgs boson, $h$, to either $WW$ or $ZZ$ is given by $s_{\beta-\alpha}$, on the other hand, the coupling of the heavier Higgs boson, $H$, is equivalent to the SM coupling multiplied by $c_{\beta-\alpha}$. Notably, the coupling between the pseudoscalar, $A$, and vector bosons is absent due to CP symmetry invariance. 
	
	%
	%
	
	\section{Numerical analysis}
	We perform a general scan over the 2HDM parameter space, as given in Table \ref{tab:mynewtable}, to explore scenarios that
	may result in a significant cross section for our processes $e^+ e^- \to H^\pm W^\mp S$ $(S=H,\ A)$. 
	We use the code \texttt{2HDMC-1.8.0} \cite{Eriksson:2009ws} to numerically scan the parameter space, check theoretical constraints, and compute the $S$, $T$ and $U$ oblique parameters. Our parameter points are further checked for consistency with experimental constraints from the direct searches for additional Higgs bosons and agreement with the Higgs signal measurements using the public codes \texttt{HiggsBounds} \cite{Bechtle:2020pkv} and \texttt{HiggsSignals} \cite{Bechtle:2020uwn}.
	Additionnaly, B-physics constraints are checked using \texttt{SuperIso} \cite{Mahmoudi:2008tp}. 
	
	
	\begin{table}[t]
		\centering
		\begin{tabular}{p{5cm} p{5cm}}
			\hline 		\hline
			Parameters & Scan range \\
			\hline 		\hline
			$m_h$ & 125.09 GeV \\[1pt] 
			$m_H$& [130, 1000] GeV\\
			$m_{A,H^{\pm}}$ & [80, 1000] GeV \\[1pt] 
			$\sin(\beta - \alpha)$&[0.97, 1] \\
			$\tan \beta$ &[0.5, 45]  \\[1pt] 
			$m_{12}^2$& [0, $10^6$] GeV$^2$ \\
			\hline 		\hline
		\end{tabular}
		\caption{2HDM type-X input parameters.}
		\label{tab:mynewtable}
	\end{table}
	
	After scrutinizing the parameter space of the model with the theoretical and experimental constraints, the resulting
	parameter points are passed to \texttt{FormCalc} \cite{Hahn:2001rv,Hahn:1998yk,Kublbeck:1990xc} to compute the corresponding cross section of each process at the $e^+e^-$ collider.

\begin{figure}[t]
		\centering
		\includegraphics[width=0.6\textwidth]{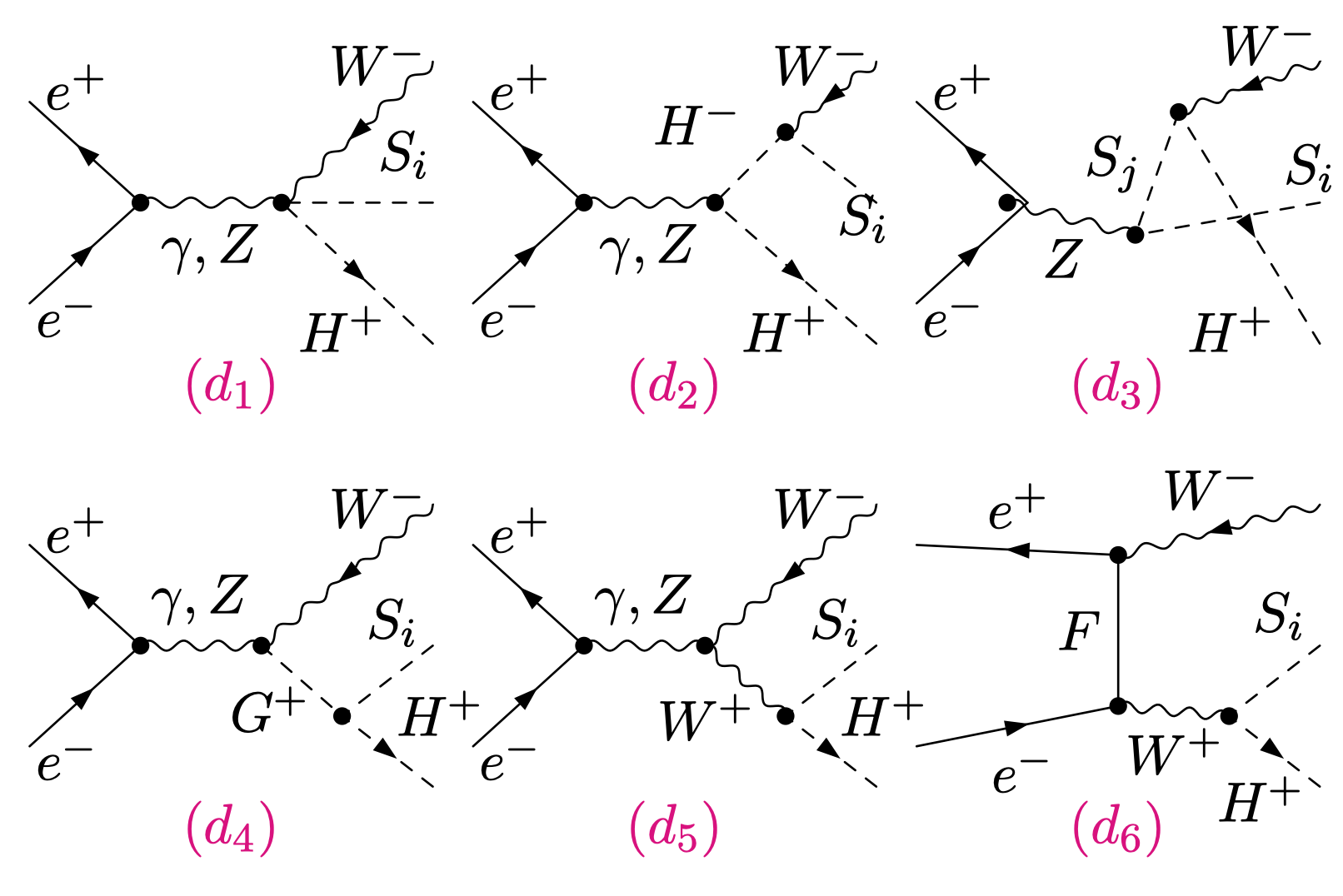}
		\caption{Tree level generic Feynman diagrams for $e^+ e^- \to H^{\pm} W^\mp S$ ($S=H,A$) are shown in $(d_{1,...,6})$. For all diagrams $S_i$ refers to $H$ or $A$.
			For $d_3$, if $S_i= H$, $S_j$ should be A, while if $S_i=A$, $S_j$ should be $H$. }
		\label{fig:fig1}
	\end{figure}

Meanwhile, the charged  Higgs boson can be produced associated with the $W^{\pm}$ boson and the neutral Higgs boson via the processes $e^+ e^- \to \{  H^{\pm}W^{\mp}H, \ H^{\pm}W^{\mp}A\}$. The relevant Feynman diagrams are shown in Fig. \ref {fig:fig1}. Most of the Feynman diagrams are mediated by the $s$-channel photon or $Z$ exchange, with the exception of the last one, $d_6$, which requires $t$-channel mediation by neutrino exchange. The couplings involved in $e^+e^- \to  H^{\pm}W^{\mp}H/A$  are pure gauge couplings and in this regard the cross section of the process is not strongly sensitive to the Yukawa textures. The Yukawa textures only enter in the width of the Higgs propagators.
After the Higgses decay into fermions, the result depends on the Yukawa type of the model. Our results here are presented for the type-X 2HDM.
It should be noted that the $e^+e^- \to  H^{\pm}W^{\mp}h$ production process is suppressed by the factor $c^2_{\beta-\alpha}$.
To ensure numerical stability of the $2\to 3$ phase space integration, we introduce the width for the scalar propagators ($d_2$ and $d_3$)  as well as for the $W$ boson propagators ($d_5$ and $d_6$).

	\begin{figure}[t]
	\centering\includegraphics[width=0.4\textwidth]{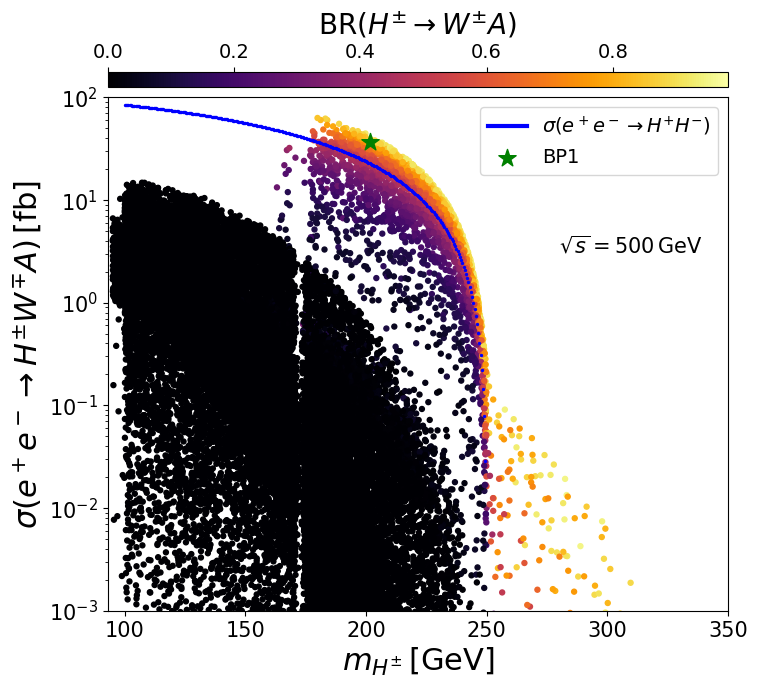}
	\includegraphics[width=0.4\textwidth]{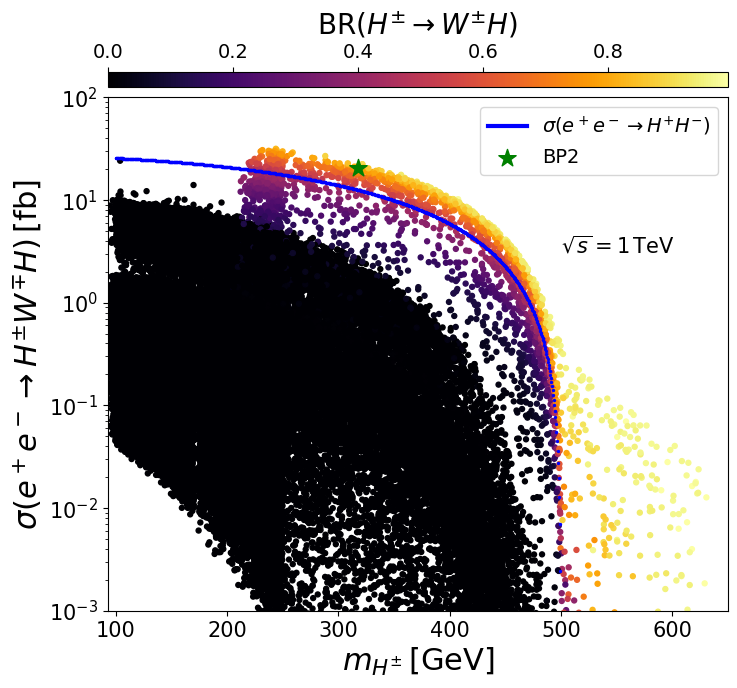}
	\includegraphics[width=0.4\textwidth]{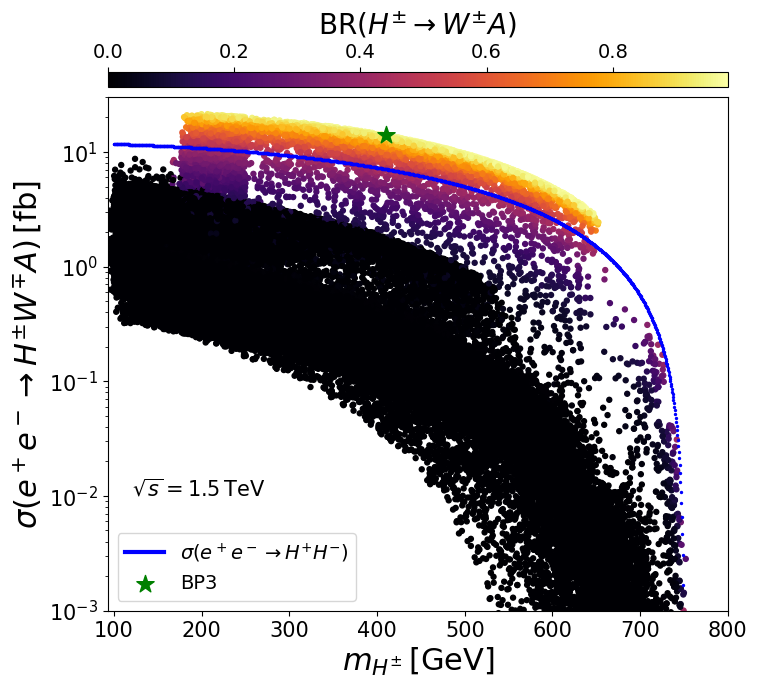}
	\caption{The cross sections $\sigma(e^+e^- \to H^{\pm}W^{\mp}S)$, at different collision energies of the ILC, as a function of $m_{H^{\pm}}$, with the blue contour lines representing $\sigma(e^+ e^- \to H^{+}H^-)$. The color bar indicates BR$(H^\pm \to W^\pm S)$. The selected Benchmark Points (BPs) are highlighted by green stars.} 
	\label{fig:fig2}
\end{figure}
	
In Fig. \ref{fig:fig2}, we show the cross sections for $\sigma(e^+ e^- \to H^{\pm}W^{\mp}H, \ H^{\pm}W^{\mp}A)$ at $\sqrt{s}=500$ GeV, 1 TeV, and 1.5 TeV. 
The process $e^+ e^- \to H^{\pm}W^{\mp}H$, shown in the top right panel, is controlled by the vertex $H^\pm W^\mp H$, which is proportional to $s_{\beta-\alpha}$ and favored by the alignment, while $e^+ e^- \to H^{\pm}W^{\mp}A$, illustrated in the top left and bottom panels, is controlled by the  $H^{\pm}H^{\mp}A$ vertex, which is independent of the $\beta-\alpha$ mixing. Only Feynman diagram $d_3$ would have $s_{\beta-\alpha}$ factor coming from $e^+ e^- \to AH^* \to H^{\pm}W^{\mp}A$. At $\sqrt{s}=500$ GeV, the process $e^+ e^- \to H^{\pm}W^{\mp}A$  can reach values as high as 60 fb for favorable mass configurations, particularly near the resonance threshold where the decay $H^\pm \to W^\pm A$ becomes kinematically allowed. For the process $e^+ e^- \to H^{\pm}W^{\mp}H $ at $\sqrt{s}=1$ TeV, which is proportional to $s^2_{\beta-\alpha}$, the cross section may reach 25 fb before the opening of $H^\pm \to W^\pm H$, and it would drop as the charged Higgs mass increases. 
Once we cross the  $H^\pm \to W^\pm H$ threshold, one can see an enhancement of the cross section compared to the pair production 
$e^+ e^- \to H^{\pm}H^{\mp}$ (blue line) coming from the resonant production originating from  $e^+e^-\to H^\pm \to W^\pm H$: $\sigma(e^+ e^- \to H^{\pm}W^{\mp}H)= \sigma(e^+ e^- \to H^{\pm}H^{\mp})\times \mathrm{BR}(H^\mp \to W^\mp H)$  from diagram $d_2$ as well as from the other Feynman diagrams . At $\sqrt{s}=1.5$ TeV,  we observe a similar pattern to that at 500 GeV. However, the available phase space for charged Higgs pair production is significantly larger.
		
\begin{figure}[t]
		\centering	
			\includegraphics[width=0.5\textwidth]{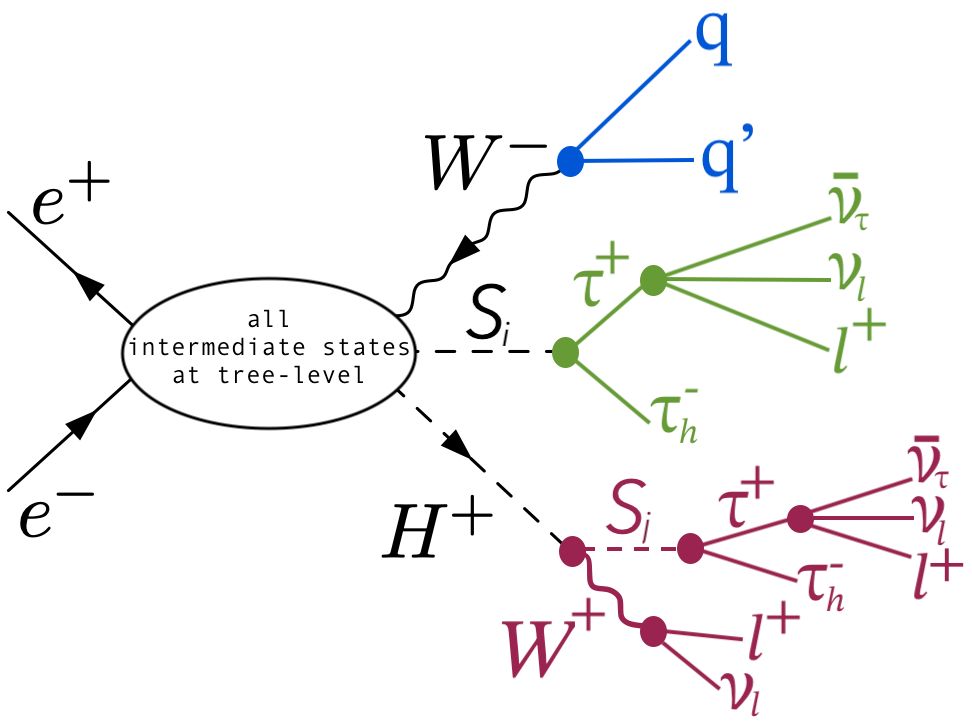}
		\caption{Diagram illustrating our two signals. Note that our analysis considers all diagrams illustrated in Fig. \ref{fig:fig1} for $e^+ e^- \to H^\pm W^\mp S~(S_i=H,A)$ production.} 
		\label{fig:fig33}
\end{figure}

In Fig. \ref{fig:fig33}, we present the first diagram, $d_1$, along with the decays chosen for our signals. However, it is important to emphasize that our analysis thoroughly considers all diagrams, including $d_{1,...,6}$, and their respective decay processes.

	\section{Signal-to-background analysis}
	In the previous section, we computed the parton-level cross sections for the proposed channels aimed at probing the  $H^\pm$ in the type-X model. While these cross sections are not small, the discovery potential depends on how effectively we can distinguish the signal from relevant backgrounds. In this section, we develop search strategies for comprehensive signal-to-background optimization, utilizing advanced tools that incorporate hard-scattering matrix elements, resonance decays, parton showers, hadronization, hadron decays, and a simplified detector response. We conduct detailed studies of the $[\tau \tau][\tau \tau]WW$ signal. For each center-of-mass energy, a specific BP is selected (see Table \ref{Bp1}). For each BP, either $H$ or $A$ is considered, but not both, as the decays $H^\pm \to W^\pm H$ and $H^\pm \to W^\pm A$ cannot simultaneously occur due to $STU$ constraints.
	Note that the  selected BPs are further checked using the new versions of \texttt{HiggsBounds} and \texttt{HiggsSignals} via \texttt{HiggsTools} \cite{Bahl:2022igd}. 
	
	\begin{table}[b]
		\setlength{\tabcolsep}{9pt}
		\renewcommand{\arraystretch}{1.6}
		\centering
		\begin{tabular}{|c|c| c| c|c| c| c|c|c|}       
			\hline  
			&signal& $m_h$  & $m_H$ & $m_A$ & $m_{H^{\pm}}$  &  $\tan \beta$  &  $\sin (\beta -\alpha)$&$m_{12}^2$  \\\hline\hline
			BP1& $H^{\pm}W^\pm A$ & 125.09 &  139.6 & 103.3  & 201.5 & 9    &0.999  &2138.14\\ 
			\hline 
			BP2&$H^{\pm}W^\mp H$  & 125.09 & 188.9 & 259.3 &318.1 & 21.5 & 0.999 &1652.63\\ \hline
			BP3& $H^{\pm}W^\mp A$ & 125.09 &  417.4 &109.1 & 410.6 & 14.8    &0.991  &11717.46\\ 
			\hline 
		\end{tabular}
		\caption{The description of our BPs. All masses are in GeV.}\label{Bp1}
	\end{table}

	Signal and background events are generated using \texttt{MadGraph5\_aMC\_v3.4.1} \cite{Alwall:2014hca}, where we used the \texttt{TauDecay} library \cite{Hagiwara:2012vz} to account for the $\tau$ decays. The generated events are then passed to \texttt{Pythia-8.2} \cite{Sjostrand:2007gs} for showering and hadronization. Jet clustering is performed using \texttt{FastJet} \cite{Cacciari:2011ma}, and fast detector simulations are conducted with \texttt{Delphes-3.4.5} \cite{deFavereau:2013fsa}, utilizing the default Delphes$\_$Card$\_$ILD detector card, which implements the anti-$k_t$ algorithm  \cite{Cacciari:2008gp} with a cone radius of $R = 0.4$. Finally, both signals and background events are analyzed using \texttt{MadAnalysis5} \cite{Conte:2013mea}.
	
	
	The $[\tau \tau] [\tau \tau]W W$ channel targets the  $H^\pm W^\mp S$ production, followed by $H^{\pm} \to W^{\pm}S$ ($S= A$, $H$):
	\begin{eqnarray} \nonumber
	e^+ e^- \to W^{\pm} H^{\mp}S &\to& W^{\pm} W^{\mp} SS\\\nonumber
	&\to& W_{qq^{'}} W_{l\nu_l}  \tau_h^{+} \tau_l^- \tau_h^+ \tau_l^- \\
	&\to& 3l +  \tau_h^+ \tau_h^+ + jj +\slashed{E}_T
	\end{eqnarray}
	where $l = e, \mu$ represents the charged lepton, and $j$ designates a light jet. The final state includes three charged leptons, two hadronic $\tau$'s, two light jets and missing transverse energy. 
	Probing the charged Higgs boson in this final state is particularly challenging, making $H^\pm$ mass reconstruction impossible.
	However, the signal significance at the final selection stage is expected to be significant, showing potential for discovery. 
	
	The main SM background contributions come from the processes $t\bar{t}$, $t\bar{t}jj$, $Zjj$ and $W^+W^-jj$. Backgrounds such as $t\bar{t}b\bar{b}$, $t\bar{t}V~(V=Z, h)$, $t\bar{t}ZZ$ and $ZZZ$ are minor.
	In each case, one of the top quarks is assumed to decay semileptonically, while the other one decays hadronically. Meanwhile, the
	boson $Z$ is assumed to decay into a $\tau^+ \tau^-$ pair and one of the $W$ boson is assumed to decay leptonically, while the other one decays into a pair of light jets. The backgrounds are as follows:
	
	\begin{itemize}
		\item	$e^+ e^- \rightarrow t\bar{t}. \ (t  \rightarrow W^+ b , \ W^+ \rightarrow l^+ \nu_{l}),\ (\bar{t} \rightarrow W^- \bar{b},\ W^- \rightarrow j j).$ 
		\item		$e^+ e^- \rightarrow t\bar{t}jj, \ (t  \rightarrow W^+ b, \ W^+\rightarrow l^+ \nu_{l}),\ (\bar{t} \rightarrow W^- \bar{b},\ W^- \rightarrow l^- \nu_{l}).$
		\item		$e^+ e^- \rightarrow Zjj, \ (Z \rightarrow \tau^+ \tau^-).$
		\item		$e^+ e^- \rightarrow W^+W^-jj, \ (W^+ \rightarrow j j),(W^- \rightarrow l^- \nu_{l}).$
	\end{itemize}
	
	For event selection, we take the following steps. The basic cuts consist of 
	\begin{equation}
	p_T^{j,l} \geq 10~\mathrm{GeV}, \ \ \ |\eta^{j,l}| < 2.5, \ \ \ \Delta R^{ll,lj,jj} \geq 0.4.\label{Bcuts}
	\end{equation}
	
	In order to assess observability, we calculate the  significance using the following expression:
	\begin{eqnarray}
	\Sigma =\sqrt{ \mathcal{L}} \frac{\sigma _s}{\sqrt{\sigma _s+\sigma _b}},
	\end{eqnarray} 
	where $\sigma_s$ and $\sigma_b$ are both the signal and background cross sections after all the cuts.
	
\begin{figure}[t]
	\begin{minipage}[b]{0.415\linewidth}
		\centering\includegraphics[width=\textwidth]{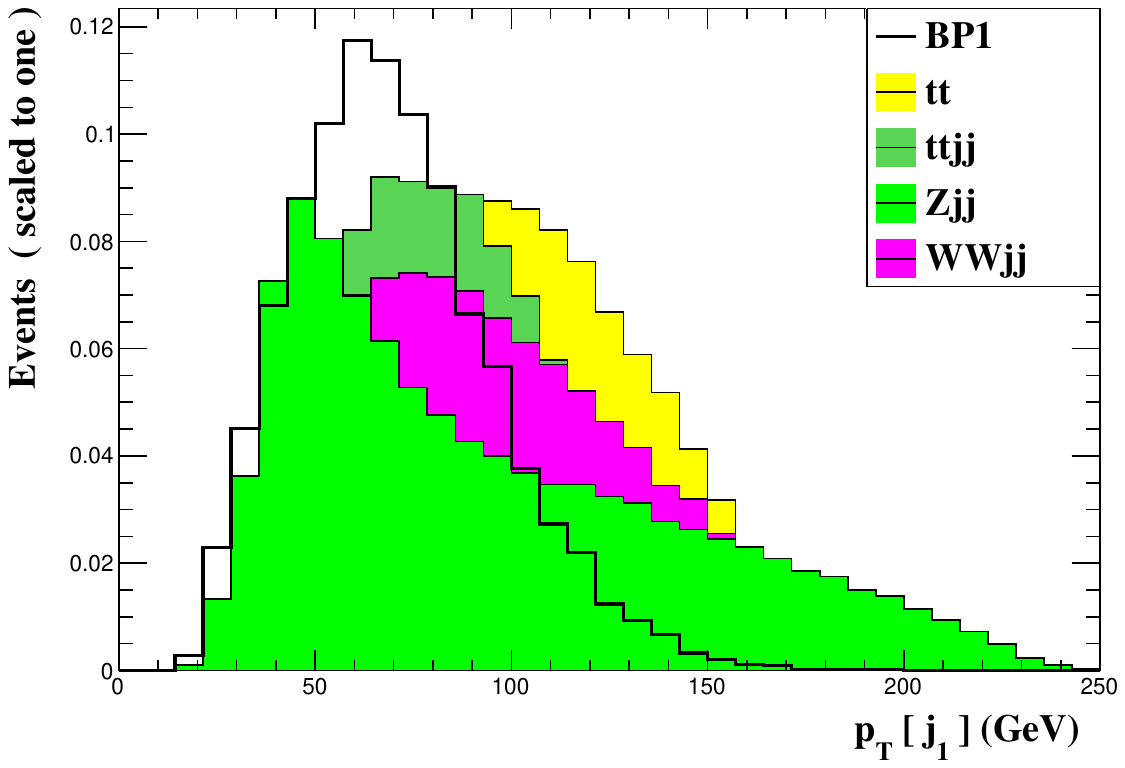}
	\end{minipage}
	\hspace{0.cm}
	\begin{minipage}[b]{0.457\linewidth}
		\centering
		\includegraphics[width=\textwidth]{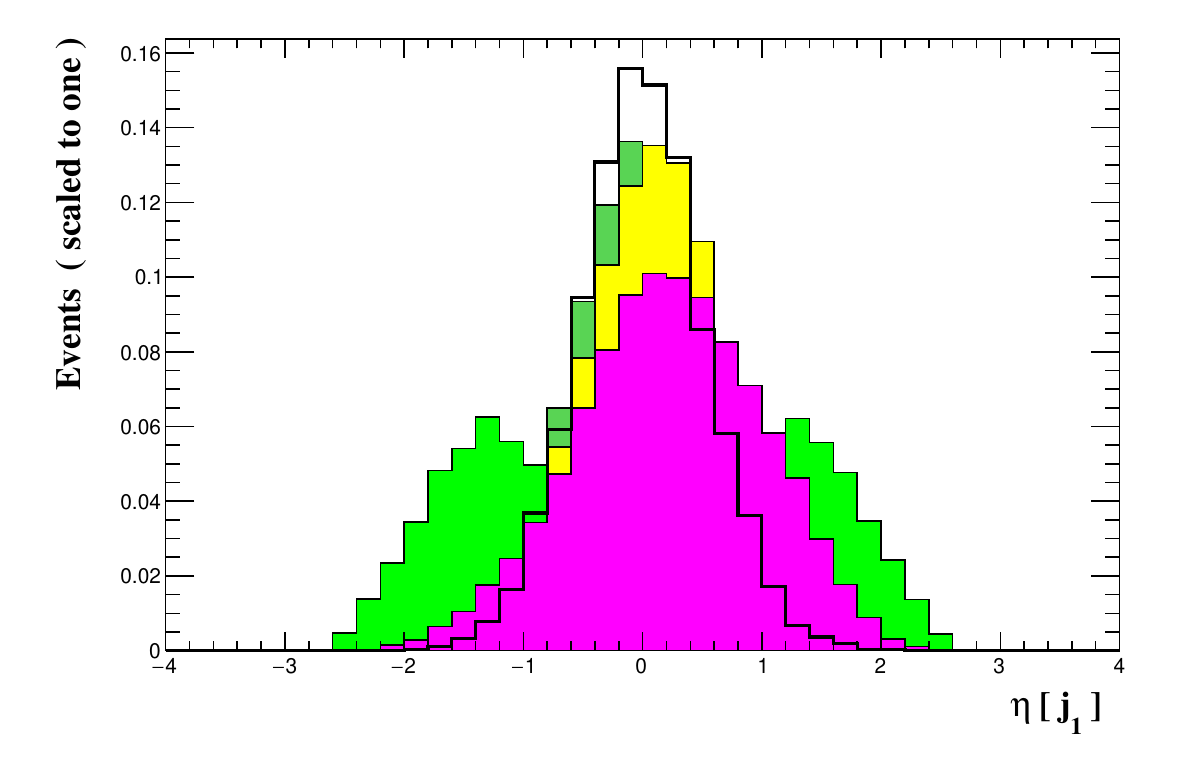}
	\end{minipage}
	\centering
	\includegraphics[width=0.45\textwidth]{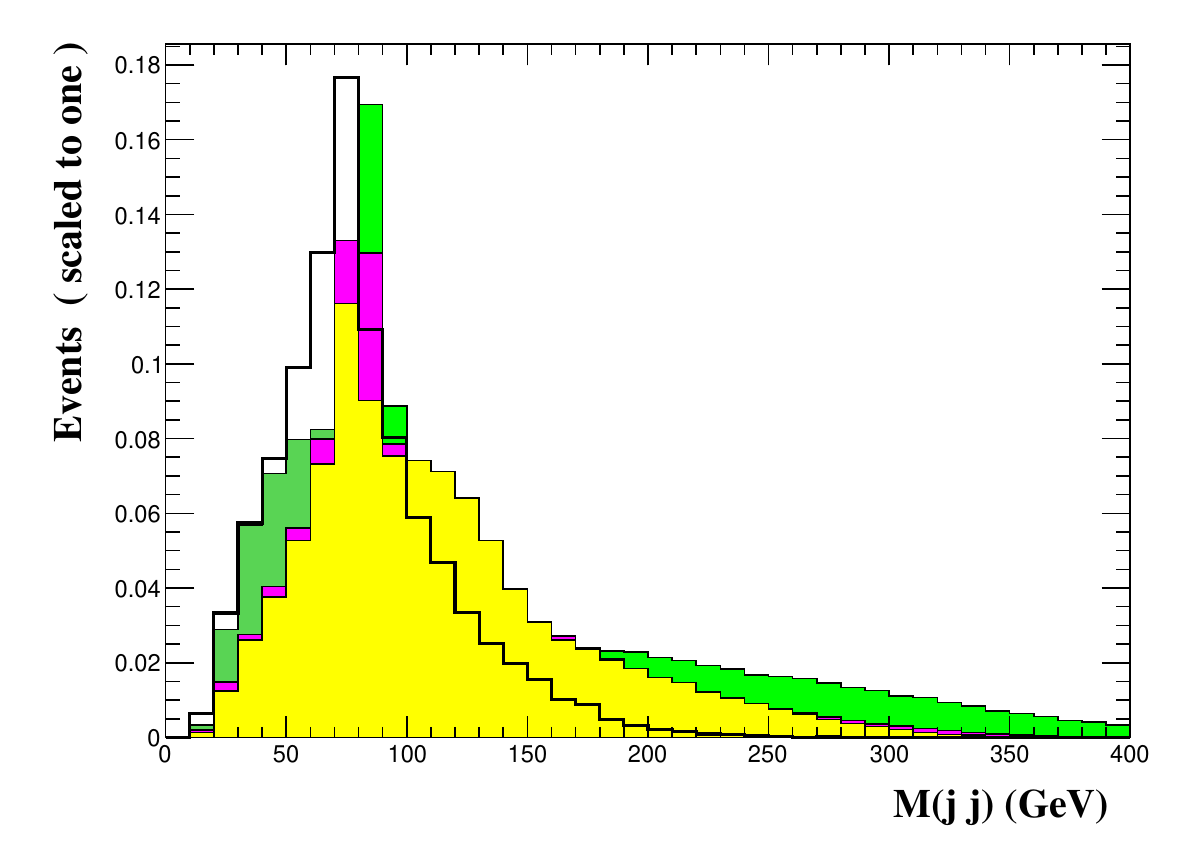}
	\caption{Selected distributions which we used in the signal to background optimization analysis: the first leading jet transverse momentum $p_T[j_1]$ (top left panel); the pseudorapidity of the first leading jet $\eta[j_1]$(top right panel); and the  mass invariant of the two leading jets $M(j \ j)$ (lower panel). The backgrounds shown here correspond to $t\bar{t}$ (yellow), $t\bar{t}jj$ (green), $Zjj$ (lime) and $WWjj$ (purple). In this canvas, we show $e^+ e^- \to H^{\pm}W^{\mp}A $ for BP1. } 
	\label{fig:fig4}
\end{figure}

	\begin{itemize}
		\centering    \item \bf{Process at $\sqrt{s} = 500~\rm{GeV}$}
	\end{itemize}

	After applying the cuts on the number of jets and $b$-jets, $N(b) \leq 1$ and $N(l) \geq 3$, which is a key discriminator between the signal and backgrounds. The signal rate remains  high, with an acceptance times efficiency $ (S \times \epsilon)$ of approximately 42$\%$. In contrast, the total background is reduced to $ (B \times \epsilon) =0.03 \%$.
	
	To divise more advanced selection strategies, the kinematic distributions of the signal and backgrounds including the first leading jet transverse momentum\footnote{The jets are labeled by descending order of $p_T$, i.e, $p_T[j_1] > p_T[j_2]$.} $p_T[j_1]$ (top-left panel), the pseudorapidity of the first leading jet $\eta[j_1]$ (top-right panel), and the invariant mass of the two leading jets $M(j \  j)$ (lower panel) are plotted in Fig.~\ref{fig:fig4}. 
			
	Based on the investigation of the kinematic distributions, we devise a search strategy summarized in Table \ref{tab:tab4}. Consequently, we impose a
	stringent criterion on the transverse momentum and the pseudorapidity of the leading jet,
	$P_T[j_1] \leq 80 $ and $-0.6 \ < \eta[j_1] < \ 0.6$, which significantly reduces a substantial portion of the background. The final selection is the invariant mass of the two leading jets, $ M(j \  j) \leq 130$ GeV, which also removes a large portion of the background events.
	At the final selection level, 50 signal events and 4 background events survive, assuming $\mathcal{L}=500$ fb$^{-1}$.

	\begin{table}[t]
		\centering
		\setlength{\tabcolsep}{5.pt}
		\renewcommand{\arraystretch}{1}
		\begin{tabular}{p{6.2cm}<{\centering}  p{1.5cm}<{\centering} p{1cm}<{\centering}p{1.2cm}<{\centering}  p{1.0cm}<{\centering} p{1.5cm}<{\centering} p{1.4cm}<{\centering} p{2.cm}<{\centering} p{2cm}<{\centering} p{2cm}<{\centering} }
			\hline\hline
			\multirow{2}{*}{Cuts$\ \ \ \ \ \ \ \ \ \ \ \ \ \ \ $}& \multicolumn{1}{c}{Signal }& \multicolumn{3}{c}{~~Backgrounds}&  \\ \cline{2-2}  \cline{4-7}
			&  $\text{BP}1$ &&$t\bar t  $& $t\bar t jj$  & $Zjj$& $W^+W^-jj$\\
			\hline\hline
			Basic cut $\ \ \ \ \ \ \ \ $      &0.63  &&74.2 &0.29&14.1 &2.99\\
			$N(b) \leq$ 1 and $N(l) \geq$ 3   &0.266  && 0.02&0.004&0.005 &0.0003 \\
			$P_T[j_1] \leq 80 $ and $-0.6 \ < \eta[j_1] < \ 0.6$ &0.133 && 0.008 &0.0018 &0.0004&9$.10^{-5}$\\
			$M(j\  j) \leq 130$ GeV &0.1&& 0.007&0.001 &0.0001 &$9.10^{-5}$\\
			Total efficiencies& $15.8\%$ &&$9.10^{-3}$& $3.10^{-1}$ &$7.10^{-6}$  &  $3.10^{-5}$ \\
			\hline\hline
		\end{tabular}
		\caption{The cut-flow chart of the cross section (in fb) counts for the $[\tau \tau][\tau \tau]WW$ signal and backgrounds, with $\sqrt{s} = 500$ GeV\label{CUTb}.}
		\label{tab:tab4}
	\end{table}
	
	\begin{table}[t]
		\setlength{\tabcolsep}{16pt}
		\renewcommand{\arraystretch}{0.8}
		\centering
		\begin{tabular}{c c c c c}       
			\hline  \hline 
			
			& &BP1& \\
			\hline  \hline 
			\hline   
			Processes $\ \ \ \ \ $   &&$e^+e^- \to H^{\pm}W^{\mp}A \to [\tau \tau] [\tau \tau]W W$ &\\
			\hline  \hline
			Luminosity$\ \ $&$\mathcal{L}$=500 fb$^{-1}$&$\mathcal{L}$=1000 fb$^{-1}$& $\mathcal{L}$=1500 fb$^{-1}$\\
			\hline  \hline 
			Significante $\Sigma$  &6.79&9.61&11.77 \\
			\hline \hline
		\end{tabular}
		\caption{Significance for our signal with $\sqrt{s}$= 500 GeV
			and $\mathcal{L}$ = 500, 1000 and 1500 fb$^{-1}$.}\label{Signi:si1}
	\end{table}
	
	In Table \ref{Signi:si1}, the statistical significance $\Sigma$ for the signal process $e^+e^- \to H^{\pm}W^{\mp}A \to [\tau \tau] [\tau \tau]W W$ at $\sqrt{s} = 500$ GeV for three different integrated luminosities: 500, 1000, and 1500 fb$^{-1}$. As expected, the significance increases with luminosity, with values of 6.79, 9.61, and 11.77, respectively. These results demonstrate the potential observability of the signal across all three considered luminosities of 500, 1000, and  1500 fb$^{-1}$, making this channel a promising probe for the considered benchmark point.
	
	\begin{itemize}
		\centering    \item \bf{Process at $\sqrt{s} = 1~\rm{TeV}$}
	\end{itemize}
	
\begin{figure}[t]
		\begin{minipage}[b]{0.45\linewidth}
			\centering\includegraphics[width=\textwidth]{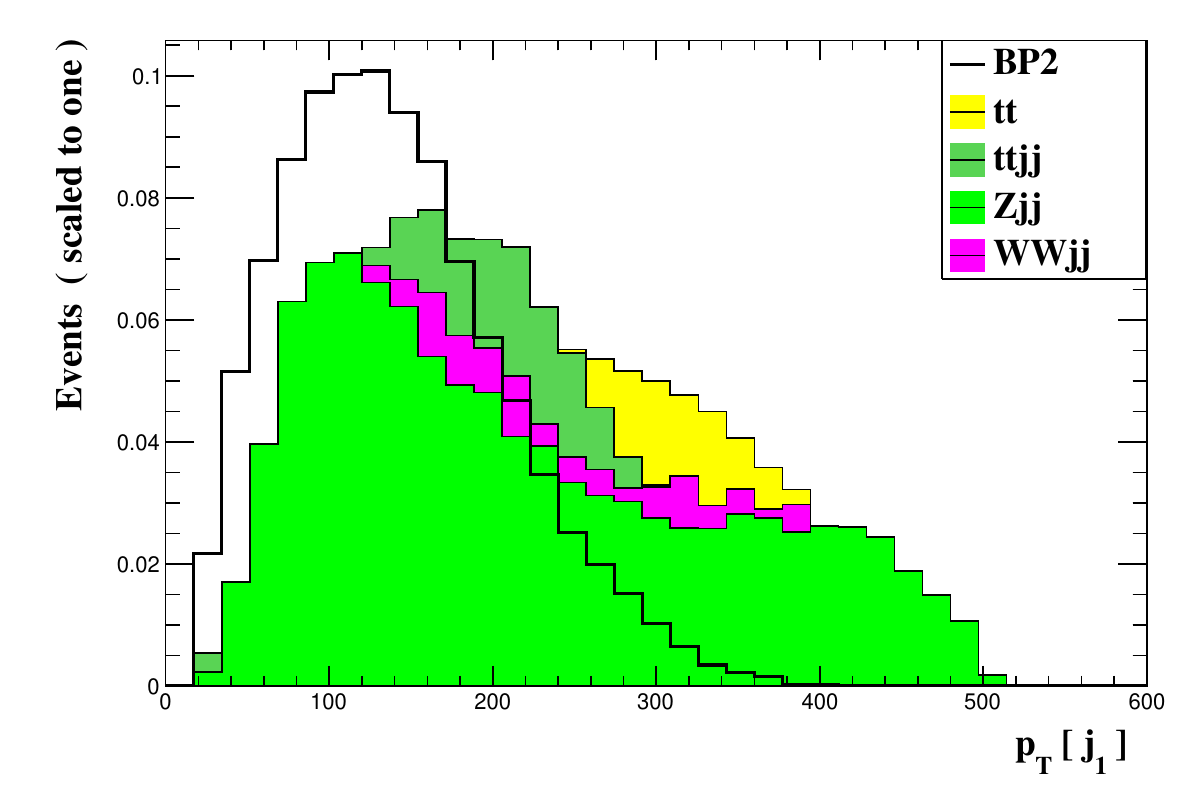}
		\end{minipage}
		\hspace{0.cm}
		\begin{minipage}[b]{0.45\linewidth}
			\centering
			\includegraphics[width=\textwidth]{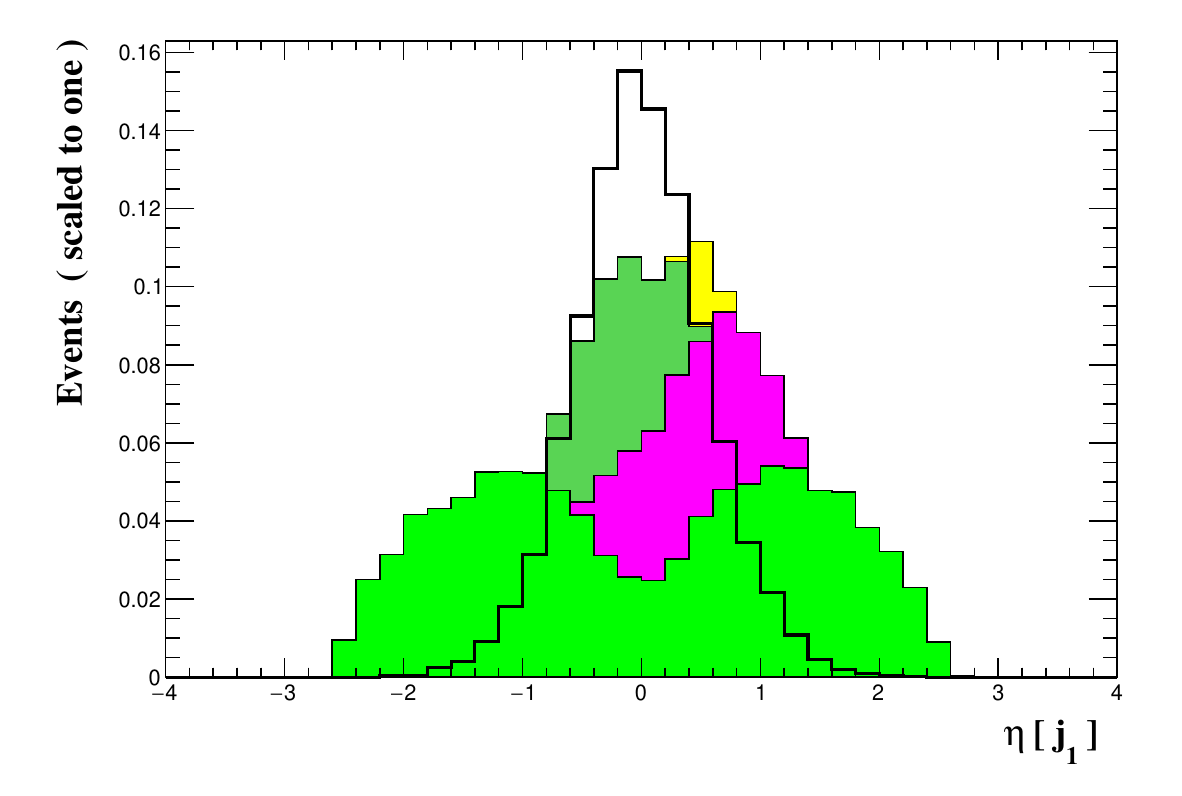}
		\end{minipage}
		\centering
		\includegraphics[width=0.45\textwidth]{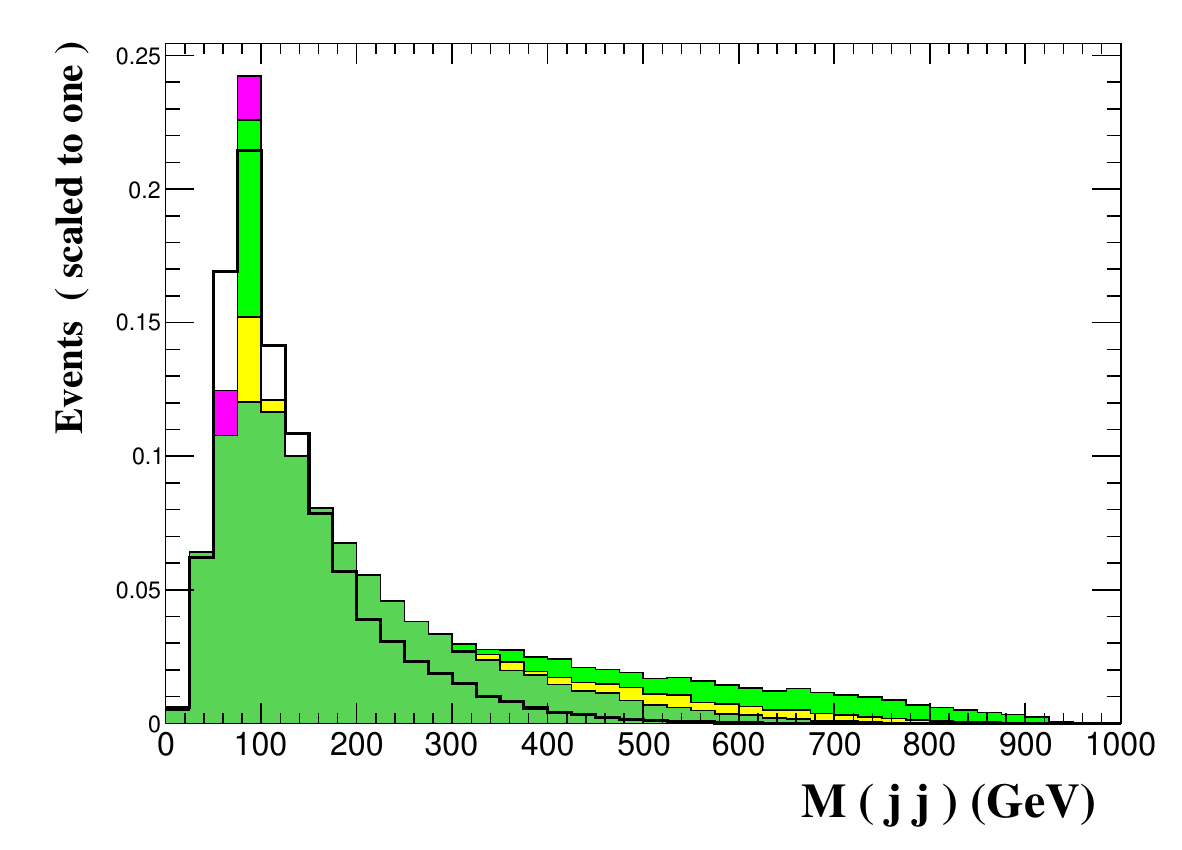}
		\caption{Selected distributions which we used in the signal to background optimization analysis: the first leading jet transverse momentum $p_T[j_1]$ (top left panel); the pseudorapidity of the first leading jet $\eta[j_1]$(top right panel); and the invariant mass of the two leading jets  $M(j\ j)$ (lower panel). The backgrounds shown here correspond to $t\bar{t}$ (yellow), $t\bar{t}jj$ (green), $Zjj$ (lime) and $WWjj$ (purple). In this canvas, we show $e^+ e^- \to H^{\pm}W^{\mp}H $ for BP2. } 
		\label{fig:fig5}
	\end{figure}

	For the signal and background events that pass the basic cuts, we calculate various kinematic distributions. In Fig. \ref{fig:fig5}, we show the distributions of the first leading jet transverse momentum $p_T[j_1]$ (top-left panel), the pseudorapidity of the first leading jet $\eta[j_1]$ (top-right panel), and the invariant mass of the two leading jets $M(j \ j)$ (lower panel)  for the signal benchmark point BP2 and various SM backgrounds at the 1 TeV ILC.

	To enhance the signal significance, we implemented a cut-flow strategy guided by the behavior of key kinematic distributions, as summarized in Table \ref{tab:tab5}.
	We impose cuts, which serve as crucial discriminators distinguishing the signal from the background, on the number of charged leptons:  $N(l) \geq 3$. With this cut applied, 48$\%$ of the signal events are retained, while the background is significantly suppressed. The first selection cut applied is $P_T[j_1] \leq 170 $ GeV and $-0.6 \ < \eta[j_1] < \ 0.6$, going together for the detection of the first leading  jet. This cut significantly reduces a substantial portion of the background, particularly eliminating the majority of $t\bar{t}$, $WWjj$ and $Zjj$ events. We investigate one of the most effective cuts, $ M(j \ j) \leq 140$ GeV, which keeps approximately 73$\%$ of the signal events while eliminating over 54$\%$ of the background events. This constitutes the final selection in our cut flow strategy, significantly enhancing the signal-to-background ratio.
	\begin{table}
		\centering
		\setlength{\tabcolsep}{5.pt}
		\renewcommand{\arraystretch}{1}
		\begin{tabular}{p{6.5cm}<{\centering}  p{1.5cm}<{\centering} p{1cm}<{\centering}p{1.2cm}<{\centering}  p{1.0cm}<{\centering} p{1.5cm}<{\centering} p{1.4cm}<{\centering} p{2.cm}<{\centering} p{1.8cm}<{\centering} p{2cm}<{\centering} }
			\hline\hline
			\multirow{2}{*}{Cuts$\ \ \ \ \ \ \ \ \ \ \ \ \ \ \ $}& \multicolumn{1}{c}{Signal }& \multicolumn{3}{c}{~~Backgrounds}&  \\ \cline{2-2}  \cline{4-7}
			&  $\text{BP}2$ &&$t\bar t  $& $t\bar t jj$  & $Zjj$& $W^+W^-jj$\\
			\hline\hline
			Basic cut $\ \ \ \ \ \ \ \ $      &0.25  &&22.5 &1.27&2.67 &2.71\\
			$N(l) \geq$ 3   &0.12  && 0.012&0.021&0.001 &0.001 \\
			$P_T[j_1] \leq 170 $ and $-0.6 \ < \eta[j_1] < \ 0.6$ &0.079 && 0.004 &0.005 &0.0&0.0002\\
			$ M(j \ j) \leq 140$ GeV &0.058 && 0.002&0.003 &0.0 &0.0\\
			Total efficiencies& $23\%$ &&8$.10^{-5}$& $0.23\%$ & ... &  ... \\
			\hline\hline
		\end{tabular}
		\caption{The cut-flow chart of the cross section (in fb) counts for the $[\tau \tau][\tau \tau]WW$ signal and backgrounds, with $\sqrt{s} = 1$ TeV. \label{CUTb}}
		\label{tab:tab5}
	\end{table}
	
	\begin{table}[t]
		\setlength{\tabcolsep}{16pt}
		\renewcommand{\arraystretch}{0.8}
		\centering
		\begin{tabular}{c c c c c}       
			\hline  \hline 
			
			& &BP2& \\
			\hline  \hline 
			\hline   
			Process $\ \ \ \ \ $   &&$e^+e^- \to H^{\pm}W^{\mp}H \to [\tau \tau] [\tau \tau]W W$ &\\
			\hline  \hline
			Luminosity$\ \ $&$\mathcal{L}$=500 fb$^{-1}$&$\mathcal{L}$=1000 fb$^{-1}$& $\mathcal{L}$=1500 fb$^{-1}$\\
			\hline  \hline 
			Significante $\Sigma$  &5.12&7.25&8.87\\
			\hline \hline
		\end{tabular}
		\caption{Significance for our signal with $\sqrt{s}$= 1 TeV
			and $\mathcal{L}$ = 500, 1000 and 1500 fb$^{-1}$.}\label{Signi:si2}
	\end{table}
	In Table \ref{Signi:si2}, the signal significance $\Sigma$ for the process $e^+e^- \to H^{\pm}W^{\mp}H \to [\tau \tau] [\tau \tau]W W$  is presented at $\sqrt{s}$= 1 TeV for three integrated luminosities: 500, 1000, and 1500 fb$^{-1}$. The corresponding significance values are 5.12, 7.25, and 8.87, respectively. This progression demonstrates the enhanced visibility of the signal with increasing luminosity and supports the viability of this channel in probing the BP2 scenario.
	
	\begin{itemize}
		\centering    \item \bf{Process at $\sqrt{s} = 1.5~\rm{TeV}$}
	\end{itemize}
	
\begin{figure}[t]
		\begin{minipage}[b]{0.45\linewidth}
			\centering\includegraphics[width=\textwidth]{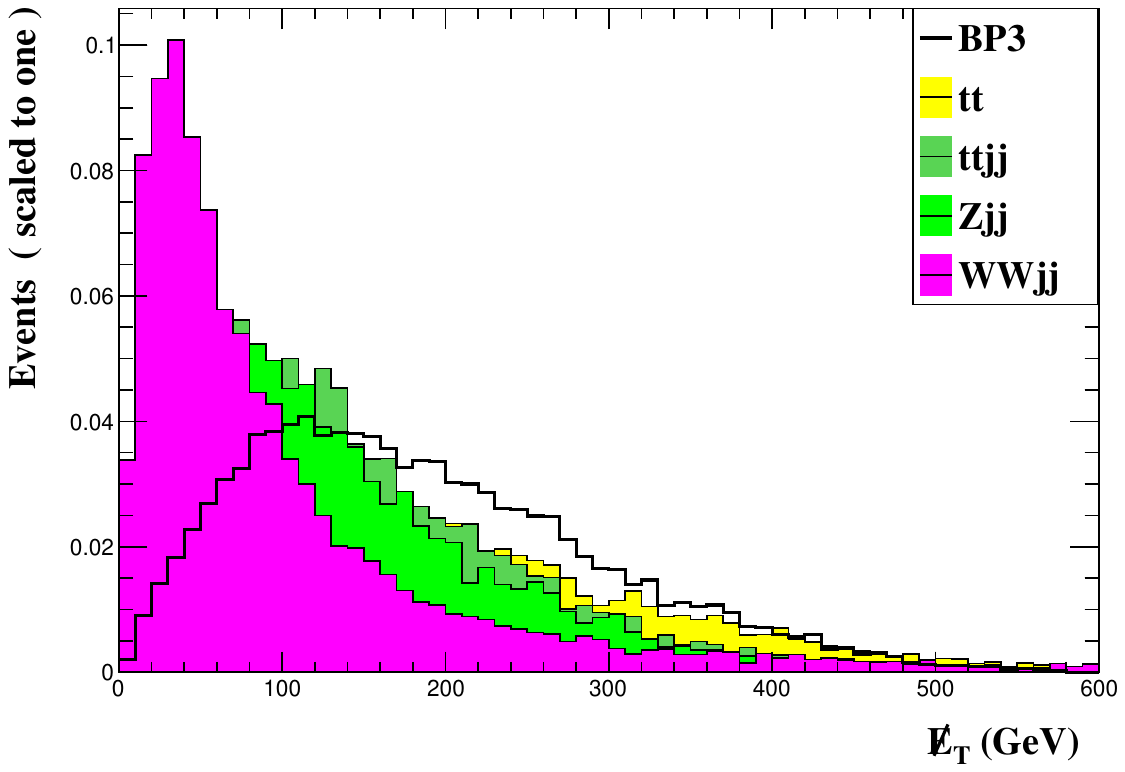}
		\end{minipage}
		\hspace{0.cm}
		\begin{minipage}[b]{0.45\linewidth}
			\centering
			\includegraphics[width=\textwidth]{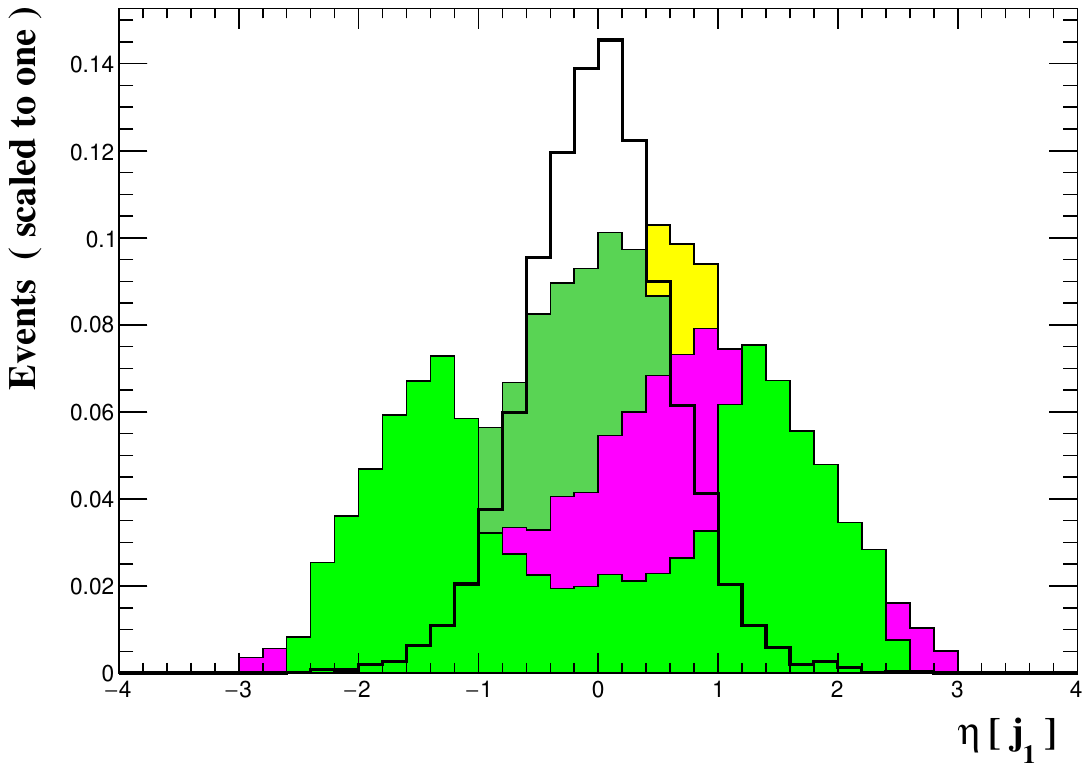}
		\end{minipage}
		\centering
		\includegraphics[width=0.45\textwidth]{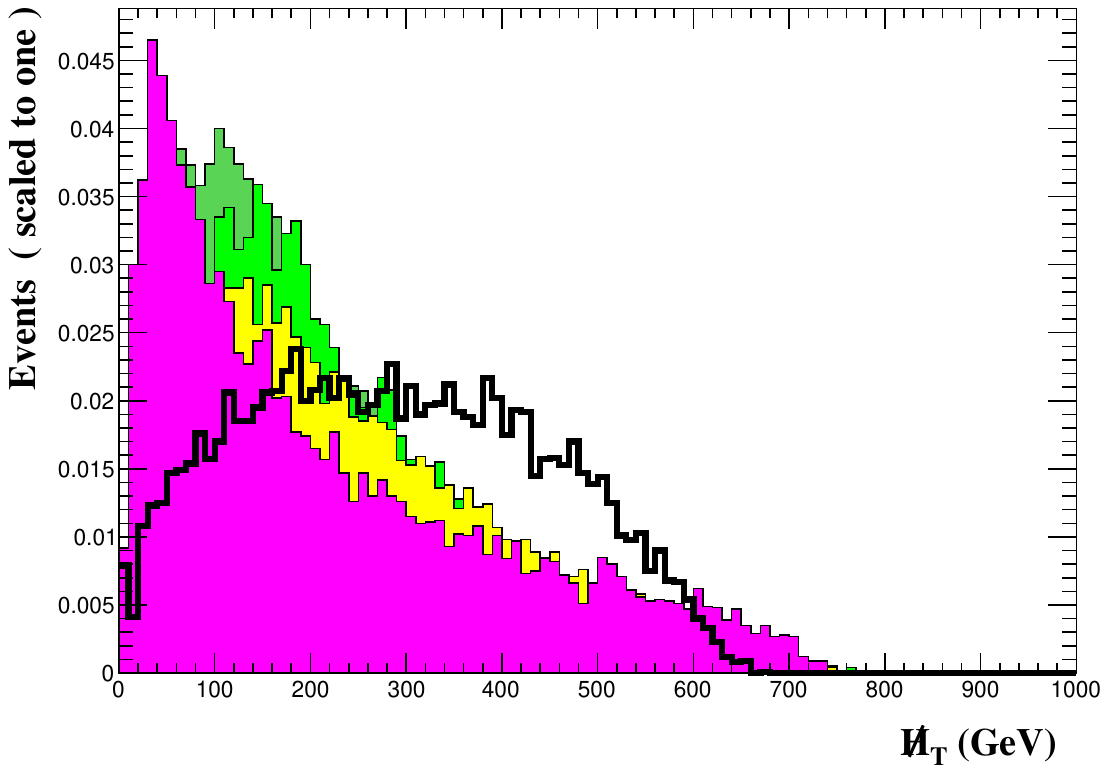}
		\caption{Selected distributions which we used in the signal-to-background optimization analysis: the missing transverse energy $\slashed{E}_T$ (top left panel); the pseudorapidity of the first leading jet $\eta[j_1]$(top right panel); and the scalar sum of jet missing transverse momentum $\slashed{H}_T$ (lower panel). The backgrounds shown here correspond to $t\bar{t}$ (yellow), $t\bar{t}jj$ (green), $Zjj$ (lime) and $WWjj$ (purple). In this canvas, we show $e^+ e^- \to H^{\pm}W^{\mp}A $ for BP3. } 
		\label{fig:fig6}
	\end{figure}

	We show in Fig. \ref{fig:fig6} the kinematic distributions of the signal and the backgrounds at $\sqrt{s}= 1.5$ TeV, specifically the missing transverse energy $\slashed{E}_T$ (top left panel), the pseudorapidity $\eta[j_1]$ (top right panel) of the first leading jet, and the scalar sum of jet missing transverse momentum $\slashed{H}_T$ (lower panel), which includes only the missing energy from the jet activity.
	
	To enhance the signal significance, we implemented a series of selection cuts guided by the behavior of the kinematic distributions. We impose restrictions on the number of $b$-jets and charged leptons, requiring $N(b) \leq $ 1 and $N(l)\geq 3$, which serves as a crucial selection to distinguish the signal from background processes. The first selection cut applied is the missing transverse energy requirement $130 \ <\slashed{E}_T < \ 500 $ GeV, which removes about 34$\%$ of the $t\bar{t}$,
	86$\%$ of the $Zjj$, 100$\%$ of $WWjj$ events, while the survival rate for the signal is more than 66$\%$.  The cut of $-0.6 \ < \eta[j_1] < \ 0.6$ 
	has a certain advantage in separating the signal from the backgrounds, especially $Zjj$. The final selection is on the  scalar sum of jet missing transverse momentum. The selection cut $\slashed{H}_T > \ 290$ GeV removes almost all the background events. Table \ref{CUT3} shows the cut-flows on the cross sections (in fb) for both the signal and the SM backgrounds at $\sqrt{s}$ =1.5 TeV. 
	\begin{table}[t]
		\centering
		\setlength{\tabcolsep}{5.pt}
		\renewcommand{\arraystretch}{1}
		\begin{tabular}{p{6.2cm}<{\centering}  p{1.5cm}<{\centering} p{1cm}<{\centering}p{1.2cm}<{\centering}  p{1.0cm}<{\centering} p{1.5cm}<{\centering} p{1.4cm}<{\centering} p{2.cm}<{\centering} p{2cm}<{\centering} p{2cm}<{\centering} }
			\hline\hline
			\multirow{2}{*}{Cuts$\ \ \ \ \ \ \ \ \ \ \ \ \ \ \ $}& \multicolumn{1}{c}{Signal }& \multicolumn{3}{c}{~~Backgrounds}&  \\ \cline{2-2}  \cline{4-7}
			&  $\text{BP}3$ &&$t\bar t  $& $t\bar t jj$  & $Zjj$& $W^+W^-jj$\\
			\hline\hline
			Basic cut $\ \ \ \ \ \ \ \ $      &0.22  &&6.9 &0.9&0.91 &3.29\\
			$N(b) \leq$ 1 and $N(l) \geq$ 3   &0.12  && 0.003&0.01&0.0007 &0.0003 \\
			$130 \ <\slashed{E}_T < \ 500 $ GeV &0.08 && 0.002 &0.004 &0.0001&0\\
			$-0.6 \ < \eta[j_1] < \ 0.6$ &0.06 && 0.05&0.002&0 &0\\
			$  \slashed{H}_T > \ 290$ GeV &0.03 && 0&0.001 &0 &0\\
			Total efficiencies& $13.6\%$ &&...& $0.1\%$ & ... & ... \\
			\hline\hline
		\end{tabular}
		\caption{The cut-flow chart of the cross section (in fb) counts for the $[\tau \tau][\tau \tau]WW$ signal and backgrounds, with $\sqrt{s} = 1.5$ TeV. \label{CUT3}}
		\label{tab:tab6}
	\end{table}		
Table \ref{Signi:si3} presents the signal significance $\Sigma$ for the process $e^+e^- \to H^{\pm}W^{\mp}A \to [\tau \tau] [\tau \tau]W W$ at $\sqrt{s}$= 1.5 TeV, evaluated for integrated luminosities of 500, 1000, and 1500 fb$^{-1}$. The significance values obtained are 3.8, 5.38, and 6.59, respectively. These results indicate a clear improvement in signal observability with increasing luminosity, confirming the sensitivity of this final state to the BP3 scenario.

\begin{table}[t]
	\setlength{\tabcolsep}{16pt}
	\renewcommand{\arraystretch}{0.8}
	\centering
	\begin{tabular}{c c c c c}       
		\hline  \hline 
		
		& &BP3& \\
		\hline  \hline 
		\hline   
		Processes $\ \ \ \ \ $   &&$e^+e^- \to H^{\pm}W^{\mp}A \to [\tau \tau] [\tau \tau]W W$ &\\
		\hline  \hline
		Luminosity$\ \ $&$\mathcal{L}$=500 fb$^{-1}$&$\mathcal{L}$=1000 fb$^{-1}$& $\mathcal{L}$=1500 fb$^{-1}$\\
		\hline  \hline 
		Significante $\Sigma$  &3.8&5.38&6.59 \\
		\hline \hline
	\end{tabular}
	\caption{Significance for our signal with $\sqrt{s}$= 1.5 TeV
		and $\mathcal{L}$ = 500, 1000 and 1500 $fb^{-1}$.}\label{Signi:si3}
\end{table}

	\section{Conclusions}
	\label{sec:conclusion}
	Within the framework of the type-X 2HDM, we have conducted a comprehensive study of the phenomenology to establish a detailed roadmap for the charged Higgs boson search. In contrast to type-II and type-Y, where the $b \to s\gamma$ constraint imposes a stringent lower bound on the charged Higgs mass ($m_{H^\pm} \gtrsim 800$ GeV \cite{Misiak:2020vlo}), type-I and type-X allows for a lighter charged Higgs. 
	However, even in type-X, the parameter space is strongly constrained by other experimental constraints, such as those from direct searches for additional Higgs bosons at LEP, Tevatron, and the LHC.  
	
	We have examined the single production of $H^\pm$ in association with a $W$ boson and a non-SM Higgs boson $H$ or $A$ at $e^+e^-$ collider. We showed that the cross section for $e^+ e^- \to H^\pm W^\mp  H$ and  $e^+ e^- \to H^\pm W^\mp A$ are promising, where the former could be useful to measure the Higgs mixing parameter $s_{\beta-\alpha}$.
	Based on these characteristics, we analyzed the $[\tau \tau] [\tau \tau]W W$ final state as a probe of charged Higgs bosons at the ILC, finding it to yield a high detection significance.
	
	In this study, we focused on the type-X 2HDM, as the decays $H/A \to \tau^+\tau^-$ are small in the type-I scenario. In contrast, $H/A \to b\bar b$ can be large in type-I, leading to the $[bb][bb]W W$ final state. A detailed study of such a final state is left for future work.
	
	\subsection*{Acknowledgments}
	KC is supported in part by the National Science and Technology Council (NSTC) of Taiwan under the grant number MoST 113-2112-M-007-041-MY3. 
	MK is supported by NSTC Grant No. 113-2639-M-002-006-ASP of Taiwan. 
	AA would like to thank the Department of Physics and CTC, National Tsing Hua University, for their hospitality during the course of this work. 
	
	\bibliographystyle{JHEP}
	\bibliography{bibliography}

\providecommand{\href}[2]{#2}\begingroup\raggedright\begin{thebibliography}{10}

\bibitem{ATLAS:2012yve}
{\bf ATLAS} Collaboration, G.~Aad et~al., {\it {Observation of a new particle
  in the search for the Standard Model Higgs boson with the ATLAS detector at
  the LHC}},  {\em Phys. Lett. B} {\bf 716} (2012) 1--29,
  [\href{http://arxiv.org/abs/1207.7214}{{\tt arXiv:1207.7214}}].

\bibitem{CMS:2012qbp}
{\bf CMS} Collaboration, S.~Chatrchyan et~al., {\it {Observation of a New Boson
  at a Mass of 125 GeV with the CMS Experiment at the LHC}},  {\em Phys. Lett.
  B} {\bf 716} (2012) 30--61, [\href{http://arxiv.org/abs/1207.7235}{{\tt
  arXiv:1207.7235}}].

\bibitem{Ait-Ouazghour:2020slc}
B.~Ait-Ouazghour and M.~Chabab, {\it {The Higgs potential in 2HDM extended with
  a real triplet scalar: A roadmap}},  {\em Int. J. Mod. Phys. A} {\bf 36}
  (2021), no.~19 2150131, [\href{http://arxiv.org/abs/2006.12233}{{\tt
  arXiv:2006.12233}}].

\bibitem{Grzadkowski:2011jks}
B.~Grzadkowski, P.~Osland, and J.~Wudka, {\it {Pragmatic Extensions of the
  Standard Model}},  {\em Acta Phys. Polon. B} {\bf 42} (2011), no.~11 2245.

\bibitem{karahan2014effects}
C.~N. Karahan and B.~Korutlu, {\it Effects of a real singlet scalar on veltman
  condition},  {\em Physics Letters B} {\bf 732} (2014) 320--324.

\bibitem{darvishi2018implication}
N.~Darvishi and M.~Krawczyk, {\it Implication of quadratic divergences
  cancellation in the two higgs doublet model},  {\em Nuclear Physics B} {\bf
  926} (2018) 167--178.

\bibitem{Ouazghour:2018mld}
B.~A. Ouazghour, A.~Arhrib, R.~Benbrik, M.~Chabab, and L.~Rahili, {\it {Theory
  and phenomenology of a two-Higgs-doublet type-II seesaw model at the LHC run
  2}},  {\em Phys. Rev. D} {\bf 100} (2019), no.~3 035031,
  [\href{http://arxiv.org/abs/1812.07719}{{\tt arXiv:1812.07719}}].

\bibitem{Kling:2020hmi}
F.~Kling, S.~Su, and W.~Su, {\it {2HDM Neutral Scalars under the LHC}},  {\em
  JHEP} {\bf 06} (2020) 163, [\href{http://arxiv.org/abs/2004.04172}{{\tt
  arXiv:2004.04172}}].

\bibitem{An:2018dwb}
F.~An et~al., {\it {Precision Higgs physics at the CEPC}},  {\em Chin. Phys. C}
  {\bf 43} (2019), no.~4 043002, [\href{http://arxiv.org/abs/1810.09037}{{\tt
  arXiv:1810.09037}}].

\bibitem{CLICPhysicsWorkingGroup:2004qvu}
{\bf CLIC Physics Working Group} Collaboration, E.~Accomando et~al., {\it
  {Physics at the CLIC multi-TeV linear collider}},  in {\em {11th
  International Conference on Hadron Spectroscopy}}, CERN Yellow Reports:
  Monographs, 6, 2004.
\newblock \href{http://arxiv.org/abs/hep-ph/0412251}{{\tt hep-ph/0412251}}.

\bibitem{Aicheler:2012bya}
{\it {A Multi-TeV Linear Collider Based on CLIC Technology}: {CLIC Conceptual
  Design Report}}, .

\bibitem{FCC:2018evy}
{\bf FCC} Collaboration, A.~Abada et~al., {\it {FCC-ee: The Lepton Collider}:
  {Future Circular Collider Conceptual Design Report Volume 2}},  {\em Eur.
  Phys. J. ST} {\bf 228} (2019), no.~2 261--623.

\bibitem{TLEPDesignStudyWorkingGroup:2013myl}
{\bf TLEP Design Study Working Group} Collaboration, M.~Bicer et~al., {\it
  {First Look at the Physics Case of TLEP}},  {\em JHEP} {\bf 01} (2014) 164,
  [\href{http://arxiv.org/abs/1308.6176}{{\tt arXiv:1308.6176}}].

\bibitem{LCCPhysicsWorkingGroup:2019fvj}
{\bf LCC Physics Working Group} Collaboration, K.~Fujii et~al., {\it {Tests of
  the Standard Model at the International Linear Collider}},
  \href{http://arxiv.org/abs/1908.11299}{{\tt arXiv:1908.11299}}.

\bibitem{Moortgat-Pick:2015lbx}
A.~Arbey et~al., {\it {Physics at the e+ e- Linear Collider}},  {\em Eur. Phys.
  J. C} {\bf 75} (2015), no.~8 371,
  [\href{http://arxiv.org/abs/1504.01726}{{\tt arXiv:1504.01726}}].

\bibitem{Han:2020pif}
T.~Han, D.~Liu, I.~Low, and X.~Wang, {\it {Electroweak Couplings of the Higgs
  Boson at a Multi-TeV Muon Collider}},
  \href{http://arxiv.org/abs/2008.12204}{{\tt arXiv:2008.12204}}.

\bibitem{Han:2020uak}
T.~Han, Z.~Liu, L.-T. Wang, and X.~Wang, {\it {WIMPs at High Energy Muon
  Colliders}},  \href{http://arxiv.org/abs/2009.11287}{{\tt arXiv:2009.11287}}.

\bibitem{Belfkir:2023vpo}
M.~Belfkir, A.~Jueid, and S.~Nasri, {\it {Boosting dark matter searches at muon
  colliders with machine learning: The mono-Higgs channel as a case study}},
  {\em PTEP} {\bf 2023} (2023), no.~12 123B03,
  [\href{http://arxiv.org/abs/2309.11241}{{\tt arXiv:2309.11241}}].

\bibitem{Jueid:2023qcf}
A.~Jueid, T.~A. Chowdhury, S.~Nasri, and S.~Saad, {\it {Probing Zee-Babu states
  at muon colliders}},  {\em Phys. Rev. D} {\bf 109} (2024), no.~7 075011,
  [\href{http://arxiv.org/abs/2306.01255}{{\tt arXiv:2306.01255}}].

\bibitem{Jana:2023ogd}
S.~Jana and S.~Klett, {\it {Muonic Force and Neutrino Non-Standard Interactions
  at Muon Colliders}},  \href{http://arxiv.org/abs/2308.07375}{{\tt
  arXiv:2308.07375}}.

\bibitem{Costantini:2020stv}
A.~Costantini, F.~De~Lillo, F.~Maltoni, L.~Mantani, O.~Mattelaer, R.~Ruiz, and
  X.~Zhao, {\it {Vector boson fusion at multi-TeV muon colliders}},  5, 2020.
\newblock \href{http://arxiv.org/abs/2005.10289}{{\tt arXiv:2005.10289}}.

\bibitem{Bandyopadhyay:2024plc}
P.~Bandyopadhyay, S.~Parashar, C.~Sen, and J.~Song, {\it {Probing Inert Triplet
  Model at a multi-TeV muon collider via vector boson fusion with forward muon
  tagging}},  {\em JHEP} {\bf 07} (2024) 253,
  [\href{http://arxiv.org/abs/2401.02697}{{\tt arXiv:2401.02697}}].

\bibitem{Han:2021udl}
T.~Han, S.~Li, S.~Su, W.~Su, and Y.~Wu, {\it {Heavy Higgs bosons in 2HDM at a
  muon collider}},  {\em Phys. Rev. D} {\bf 104} (2021), no.~5 055029,
  [\href{http://arxiv.org/abs/2102.08386}{{\tt arXiv:2102.08386}}].

\bibitem{Akeroyd:2016ymd}
A.~G. Akeroyd et~al., {\it {Prospects for charged Higgs searches at the LHC}},
  {\em Eur. Phys. J. C} {\bf 77} (2017), no.~5 276,
  [\href{http://arxiv.org/abs/1607.01320}{{\tt arXiv:1607.01320}}].

\bibitem{Barger:1993th}
V.~D. Barger, R.~J.~N. Phillips, and D.~P. Roy, {\it {Heavy charged Higgs
  signals at the LHC}},  {\em Phys. Lett. B} {\bf 324} (1994) 236--240,
  [\href{http://arxiv.org/abs/hep-ph/9311372}{{\tt hep-ph/9311372}}].

\bibitem{ILC:2013jhg}
{\bf ILC} Collaboration, {\it {The International Linear Collider Technical
  Design Report - Volume 2: Physics}},
  \href{http://arxiv.org/abs/1306.6352}{{\tt arXiv:1306.6352}}.

\bibitem{ECFADESYLCPhysicsWorkingGroup:2001igx}
{\bf ECFA/DESY LC Physics Working Group} Collaboration, J.~A. Aguilar-Saavedra
  et~al., {\it {TESLA: The Superconducting electron positron linear collider
  with an integrated x-ray laser laboratory. Technical design report. Part 3.
  Physics at an e+ e- linear collider}},
  \href{http://arxiv.org/abs/hep-ph/0106315}{{\tt hep-ph/0106315}}.

\bibitem{Ouazghour:2024twx}
B.~A. Ouazghour, A.~Arhrib, K.~Cheung, E.-s. Ghourmin, and L.~Rahili, {\it
  {Associated charged Higgs boson production within the 2HDM: e-e+ versus
  \ensuremath{\mu}-\ensuremath{\mu}+ colliders}},  {\em Phys. Rev. D} {\bf 110}
  (2024), no.~9 095026, [\href{http://arxiv.org/abs/2408.13952}{{\tt
  arXiv:2408.13952}}].

\bibitem{Komamiya:1988rs}
S.~Komamiya, {\it {Searching for Charged Higgs Bosons at O (1/2-tev to 1-tev)
  $e^+ e^-$ Colliders}},  {\em Phys. Rev. D} {\bf 38} (1988) 2158.

\bibitem{Kanemura:2000cw}
S.~Kanemura, S.~Moretti, and K.~Odagiri, {\it {Single charged Higgs boson
  production at next generation linear colliders}},  {\em JHEP} {\bf 02} (2001)
  011, [\href{http://arxiv.org/abs/hep-ph/0012030}{{\tt hep-ph/0012030}}].

\bibitem{Brignole:1991fw}
A.~Brignole, J.~R. Ellis, .~J.~F. Gunion, M.~Guzzo, F.~I. Olness, G.~Ridolfi,
  L.~Roszkowski, and F.~Zwirner, {\it {Higgs bosons in the minimal
  supersymmetric extension of the Standard Model}},  in {\em {Workshop on e+ e-
  Collisions at 500 GeV: the Physics Potential}}, 2, 1991.

\bibitem{Arhrib:1999rg}
A.~Arhrib, M.~Capdequi~Peyranere, W.~Hollik, and G.~Moultaka, {\it {Associated
  $H^- W^+$ production in high-energy $e^+ e^-$ collisions}},  {\em Nucl. Phys.
  B} {\bf 581} (2000) 34--60, [\href{http://arxiv.org/abs/hep-ph/9912527}{{\tt
  hep-ph/9912527}}]. [Erratum: Nucl.Phys. 2004, 400--401 (2004)].

\bibitem{Kanemura:1999tg}
S.~Kanemura, {\it {Possible enhancement of the e+ e- ---\ensuremath{>} H+- W-+
  cross-section in the two Higgs doublet model}},  {\em Eur. Phys. J. C} {\bf
  17} (2000) 473--486, [\href{http://arxiv.org/abs/hep-ph/9911541}{{\tt
  hep-ph/9911541}}].

\bibitem{Ouazghour:2023plc}
B.~A. Ouazghour, A.~Arhrib, K.~Cheung, E.-s. Ghourmin, and L.~Rahili, {\it
  {Charged Higgs production at the Muon Collider in the 2HDM}},
  \href{http://arxiv.org/abs/2308.15664}{{\tt arXiv:2308.15664}}.

\bibitem{DELPHI:2003eid}
{\bf DELPHI} Collaboration, J.~Abdallah et~al., {\it {Search for charged Higgs
  bosons at LEP in general two Higgs doublet models}},  {\em Eur. Phys. J. C}
  {\bf 34} (2004) 399--418, [\href{http://arxiv.org/abs/hep-ex/0404012}{{\tt
  hep-ex/0404012}}].

\bibitem{ALEPH:2013htx}
{\bf ALEPH, DELPHI, L3, OPAL, LEP} Collaboration, G.~Abbiendi et~al., {\it
  {Search for Charged Higgs bosons: Combined Results Using LEP Data}},  {\em
  Eur. Phys. J. C} {\bf 73} (2013) 2463,
  [\href{http://arxiv.org/abs/1301.6065}{{\tt arXiv:1301.6065}}].

\bibitem{CDF:2005acr}
{\bf CDF} Collaboration, A.~Abulencia et~al., {\it {Search for charged Higgs
  bosons from top quark decays in $p\bar{p}$ collisions at $\sqrt{s} =$
  1.96-TeV.}},  {\em Phys. Rev. Lett.} {\bf 96} (2006) 042003,
  [\href{http://arxiv.org/abs/hep-ex/0510065}{{\tt hep-ex/0510065}}].

\bibitem{D0:2009hbc}
{\bf D0} Collaboration, V.~M. Abazov et~al., {\it {Search for charged Higgs
  bosons in decays of top quarks}},  {\em Phys. Rev. D} {\bf 80} (2009) 051107,
  [\href{http://arxiv.org/abs/0906.5326}{{\tt arXiv:0906.5326}}].

\bibitem{ATLAS:2012nhc}
{\bf ATLAS} Collaboration, G.~Aad et~al., {\it {Search for charged Higgs bosons
  decaying via $H^{+} \to \tau \nu$ in top quark pair events using $pp$
  collision data at $\sqrt{s}=7$ TeV with the ATLAS detector}},  {\em JHEP}
  {\bf 06} (2012) 039, [\href{http://arxiv.org/abs/1204.2760}{{\tt
  arXiv:1204.2760}}].

\bibitem{ATLAS:2012tny}
{\bf ATLAS} Collaboration, G.~Aad et~al., {\it {Search for charged Higgs bosons
  through the violation of lepton universality in $t\bar{t}$ events using $pp$
  collision data at $\sqrt{s}=7$ TeV with the ATLAS experiment}},  {\em JHEP}
  {\bf 03} (2013) 076, [\href{http://arxiv.org/abs/1212.3572}{{\tt
  arXiv:1212.3572}}].

\bibitem{ATLAS:2016avi}
{\bf ATLAS} Collaboration, M.~Aaboud et~al., {\it {Search for charged Higgs
  bosons produced in association with a top quark and decaying via $H^{\pm}
  \rightarrow \tau\nu$ using $pp$ collision data recorded at $\sqrt{s} = 13$
  TeV by the ATLAS detector}},  {\em Phys. Lett. B} {\bf 759} (2016) 555--574,
  [\href{http://arxiv.org/abs/1603.09203}{{\tt arXiv:1603.09203}}].

\bibitem{ATLAS:2014otc}
{\bf ATLAS} Collaboration, G.~Aad et~al., {\it {Search for charged Higgs bosons
  decaying via $H^{\pm} \rightarrow \tau^{\pm}\nu$ in fully hadronic final
  states using $pp$ collision data at $\sqrt{s} = 8$ TeV with the ATLAS
  detector}},  {\em JHEP} {\bf 03} (2015) 088,
  [\href{http://arxiv.org/abs/1412.6663}{{\tt arXiv:1412.6663}}].

\bibitem{ATLAS:2018gfm}
{\bf ATLAS} Collaboration, M.~Aaboud et~al., {\it {Search for charged Higgs
  bosons decaying via $H^{\pm} \to \tau^{\pm}\nu_{\tau}$ in the $\tau$+jets and
  $\tau$+lepton final states with 36 fb$^{-1}$ of $pp$ collision data recorded
  at $\sqrt{s} = 13$ TeV with the ATLAS experiment}},  {\em JHEP} {\bf 09}
  (2018) 139, [\href{http://arxiv.org/abs/1807.07915}{{\tt arXiv:1807.07915}}].

\bibitem{ATLAS:2024hya}
{\bf ATLAS} Collaboration, G.~Aad et~al., {\it {Search for charged Higgs bosons
  produced in top-quark decays or in association with top quarks and decaying
  via
  H\ensuremath{\pm}\textrightarrow{}\ensuremath{\tau}\ensuremath{\pm}\ensuremath{\nu}\ensuremath{\tau}
  in 13~TeV pp collisions with the ATLAS detector}},  {\em Phys. Rev. D} {\bf
  111} (2025), no.~7 072006, [\href{http://arxiv.org/abs/2412.17584}{{\tt
  arXiv:2412.17584}}].

\bibitem{CMS:2012fgz}
{\bf CMS} Collaboration, S.~Chatrchyan et~al., {\it {Search for a Light Charged
  Higgs Boson in Top Quark Decays in $pp$ Collisions at $\sqrt{s}=7$ TeV}},
  {\em JHEP} {\bf 07} (2012) 143, [\href{http://arxiv.org/abs/1205.5736}{{\tt
  arXiv:1205.5736}}].

\bibitem{CMS:2014cdp}
C.~Collaboration, {\it Search for a charged higgs boson in pp collisions at $
  \sqrt{s} =8$ tev},  {\em JHEP} {\bf 11} (2015) 018,
  [\href{http://arxiv.org/abs/1508.07774}{{\tt arXiv:1508.07774}}].

\bibitem{CMS:2014pea}
C.~Collaboration, {\it {Search for neutral MSSM Higgs bosons decaying to a pair
  of tau leptons in pp collisions}},  {\em JHEP} {\bf 10} (2014) 160,
  [\href{http://arxiv.org/abs/1408.3316}{{\tt arXiv:1408.3316}}].

\bibitem{CMS:2015lsf}
{\bf CMS} Collaboration, V.~Khachatryan et~al., {\it {Search for a charged
  Higgs boson in pp collisions at $ \sqrt{s}=8 $ TeV}},  {\em JHEP} {\bf 11}
  (2015) 018, [\href{http://arxiv.org/abs/1508.07774}{{\tt arXiv:1508.07774}}].

\bibitem{CMS:2019bfg}
{\bf CMS} Collaboration, A.~M. Sirunyan et~al., {\it {Search for charged Higgs
  bosons in the H$^{\pm}$ $\to$ $\tau^{\pm}\nu_\tau$ decay channel in
  proton-proton collisions at $\sqrt{s} =$ 13 TeV}},  {\em JHEP} {\bf 07}
  (2019) 142, [\href{http://arxiv.org/abs/1903.04560}{{\tt arXiv:1903.04560}}].

\bibitem{CMS:2018ect}
{\bf CMS} Collaboration, {\it {Search for charged Higgs bosons with the
  H$^{\pm} \to \tau^{\pm}\nu_\tau$ decay channel in proton-proton collisions at
  $\sqrt{s}=13~\mathrm{TeV}$}}, .

\bibitem{ATLAS:2013uxj}
{\bf ATLAS} Collaboration, G.~Aad et~al., {\it {Search for a light charged
  Higgs boson in the decay channel $H^+ \to c\bar{s}$ in $t\bar{t}$ events
  using pp collisions at $\sqrt{s}$ = 7 TeV with the ATLAS detector}},  {\em
  Eur. Phys. J. C} {\bf 73} (2013), no.~6 2465,
  [\href{http://arxiv.org/abs/1302.3694}{{\tt arXiv:1302.3694}}].

\bibitem{CMS:2015yvc}
{\bf CMS} Collaboration, V.~Khachatryan et~al., {\it {Search for a light
  charged Higgs boson decaying to $ \mathrm{c}\overline{\mathrm{s}} $ in pp
  collisions at $ \sqrt{s}=8 $ TeV}},  {\em JHEP} {\bf 12} (2015) 178,
  [\href{http://arxiv.org/abs/1510.04252}{{\tt arXiv:1510.04252}}].

\bibitem{CMS:2020osd}
{\bf CMS} Collaboration, A.~M. Sirunyan et~al., {\it {Search for a light
  charged Higgs boson in the H$^\pm$ $\to $ cs channel in proton-proton
  collisions at $\sqrt{s} =$ 13 TeV}},  {\em Phys. Rev. D} {\bf 102} (2020),
  no.~7 072001, [\href{http://arxiv.org/abs/2005.08900}{{\tt
  arXiv:2005.08900}}].

\bibitem{ATLAS:2024oqu}
{\bf ATLAS} Collaboration, G.~Aad et~al., {\it {Search for a light charged
  Higgs boson in $t \rightarrow H^{\pm } b$ decays, with $H^{\pm } \rightarrow
  cs$, in $pp$ collisions at $\sqrt{s}={13}\hbox { TeV}$ with the ATLAS
  detector}},  {\em Eur. Phys. J. C} {\bf 85} (2025), no.~2 153,
  [\href{http://arxiv.org/abs/2407.10096}{{\tt arXiv:2407.10096}}].

\bibitem{ATLAS:2023bzb}
{\bf ATLAS} Collaboration, G.~Aad et~al., {\it {Search for a light charged
  Higgs boson in $t \rightarrow H^{\pm}b$ decays, with $H^{\pm} \rightarrow
  cb$, in the lepton+jets final state in proton-proton collisions at
  $\sqrt{s}=13$ TeV with the ATLAS detector}},  {\em JHEP} {\bf 09} (2023) 004,
  [\href{http://arxiv.org/abs/2302.11739}{{\tt arXiv:2302.11739}}].

\bibitem{Arhrib:2024sfg}
A.~Arhrib, M.~Krab, and S.~Semlali, {\it {Accommodating the LHC charged Higgs
  boson excess at 130 GeV in the general two-Higgs doublet model}},  {\em J.
  Phys. G} {\bf 51} (2024), no.~11 115003,
  [\href{http://arxiv.org/abs/2402.03195}{{\tt arXiv:2402.03195}}].

\bibitem{CMS:2018dzl}
{\bf CMS} Collaboration, A.~M. Sirunyan et~al., {\it {Search for a charged
  Higgs boson decaying to charm and bottom quarks in proton-proton collisions
  at $ \sqrt{s}=8 $ TeV}},  {\em JHEP} {\bf 11} (2018) 115,
  [\href{http://arxiv.org/abs/1808.06575}{{\tt arXiv:1808.06575}}].

\bibitem{ATLAS:2015nkq}
{\bf ATLAS} Collaboration, G.~Aad et~al., {\it {Search for charged Higgs bosons
  in the $H^{\pm} \rightarrow tb$ decay channel in $pp$ collisions at
  $\sqrt{s}=8 $ TeV using the ATLAS detector}},  {\em JHEP} {\bf 03} (2016)
  127, [\href{http://arxiv.org/abs/1512.03704}{{\tt arXiv:1512.03704}}].

\bibitem{ATLAS:2018ntn}
{\bf ATLAS} Collaboration, M.~Aaboud et~al., {\it {Search for charged Higgs
  bosons decaying into top and bottom quarks at $\sqrt{s}$ = 13 TeV with the
  ATLAS detector}},  {\em JHEP} {\bf 11} (2018) 085,
  [\href{http://arxiv.org/abs/1808.03599}{{\tt arXiv:1808.03599}}].

\bibitem{ATLAS:2021upq}
{\bf ATLAS} Collaboration, G.~Aad et~al., {\it {Search for charged Higgs bosons
  decaying into a top quark and a bottom quark at $ \sqrt{\mathrm{s}} $ = 13
  TeV with the ATLAS detector}},  {\em JHEP} {\bf 06} (2021) 145,
  [\href{http://arxiv.org/abs/2102.10076}{{\tt arXiv:2102.10076}}].

\bibitem{CMS:2019rlz}
{\bf CMS} Collaboration, A.~M. Sirunyan et~al., {\it {Search for a charged
  Higgs boson decaying into top and bottom quarks in events with electrons or
  muons in proton-proton collisions at $ \sqrt{\mathrm{s}} $ = 13 TeV}},  {\em
  JHEP} {\bf 01} (2020) 096, [\href{http://arxiv.org/abs/1908.09206}{{\tt
  arXiv:1908.09206}}].

\bibitem{CMS:2020imj}
{\bf CMS} Collaboration, A.~M. Sirunyan et~al., {\it {Search for charged Higgs
  bosons decaying into a top and a bottom quark in the all-jet final state of
  pp collisions at $ \sqrt{s} $ = 13 TeV}},  {\em JHEP} {\bf 07} (2020) 126,
  [\href{http://arxiv.org/abs/2001.07763}{{\tt arXiv:2001.07763}}].

\bibitem{Arhrib:2016wpw}
A.~Arhrib, R.~Benbrik, and S.~Moretti, {\it {Bosonic Decays of Charged Higgs
  Bosons in a 2HDM Type-I}},  {\em Eur. Phys. J. C} {\bf 77} (2017), no.~9 621,
  [\href{http://arxiv.org/abs/1607.02402}{{\tt arXiv:1607.02402}}].

\bibitem{Arhrib:2020tqk}
A.~Arhrib, R.~Benbrik, H.~Harouiz, S.~Moretti, Y.~Wang, and Q.-S. Yan, {\it
  {Implications of a light charged Higgs boson at the LHC run III in the
  2HDM}},  {\em Phys. Rev. D} {\bf 102} (2020), no.~11 115040,
  [\href{http://arxiv.org/abs/2003.11108}{{\tt arXiv:2003.11108}}].

\bibitem{Bahl:2021str}
H.~Bahl, T.~Stefaniak, and J.~Wittbrodt, {\it {The forgotten channels: charged
  Higgs boson decays to a W$^{±}$ and a non-SM-like Higgs boson}},  {\em JHEP}
  {\bf 06} (2021) 183, [\href{http://arxiv.org/abs/2103.07484}{{\tt
  arXiv:2103.07484}}].

\bibitem{Arhrib:2021xmc}
A.~Arhrib, R.~Benbrik, M.~Krab, B.~Manaut, S.~Moretti, Y.~Wang, and Q.-S. Yan,
  {\it {New discovery modes for a light charged Higgs boson at the LHC}},  {\em
  JHEP} {\bf 10} (2021) 073, [\href{http://arxiv.org/abs/2106.13656}{{\tt
  arXiv:2106.13656}}].

\bibitem{Arhrib:2021yqf}
A.~Arhrib, R.~Benbrik, M.~Krab, B.~Manaut, S.~Moretti, Y.~Wang, and Q.-S. Yan,
  {\it {New Light H\ensuremath{\pm} Discovery Channels at the LHC}},  {\em
  Symmetry} {\bf 13} (2021), no.~12 2319,
  [\href{http://arxiv.org/abs/2110.04823}{{\tt arXiv:2110.04823}}].

\bibitem{Mondal:2021bxa}
T.~Mondal and P.~Sanyal, {\it {Same sign trilepton as signature of charged
  Higgs in two Higgs doublet model}},  {\em JHEP} {\bf 05} (2022) 040,
  [\href{http://arxiv.org/abs/2109.05682}{{\tt arXiv:2109.05682}}].

\bibitem{Arhrib:2022inj}
A.~Arhrib, R.~Benbrik, M.~Krab, B.~Manaut, S.~Moretti, Y.~Wang, and Q.~S. Yan,
  {\it {Light charged Higgs boson in $H^\pm h$ associated production at the
  LHC}},  in {\em {1st Pan-African Astro-Particle and Collider Physics
  Workshop}}, 5, 2022.
\newblock \href{http://arxiv.org/abs/2205.14274}{{\tt arXiv:2205.14274}}.

\bibitem{Krab:2022lih}
M.~Krab, M.~Ouchemhou, A.~Arhrib, R.~Benbrik, B.~Manaut, and Q.-S. Yan, {\it
  {Single charged Higgs boson production at the LHC}},  {\em Phys. Lett. B}
  {\bf 839} (2023) 137705, [\href{http://arxiv.org/abs/2210.09416}{{\tt
  arXiv:2210.09416}}].

\bibitem{CMS:2019idx}
{\bf CMS} Collaboration, A.~M. Sirunyan et~al., {\it {Search for a light
  charged Higgs boson decaying to a W boson and a CP-odd Higgs boson in final
  states with e$\mu\mu$ or $\mu\mu\mu$ in proton-proton collisions at $\sqrt{s}
  =$ 13 TeV}},  {\em Phys. Rev. Lett.} {\bf 123} (2019), no.~13 131802,
  [\href{http://arxiv.org/abs/1905.07453}{{\tt arXiv:1905.07453}}].

\bibitem{ATLAS:2021xhq}
{\bf ATLAS} Collaboration, {\it {Search for $H^{\pm} \rightarrow W^{\pm}A
  \rightarrow W^{\pm}\mu\mu$ in $pp \rightarrow t\overline{t}$ events using an
  $e\mu\mu$ signature with the ATLAS detector at $\sqrt{s}=13$ TeV}}, .

\bibitem{CMS:2022jqc}
{\bf CMS} Collaboration, A.~Tumasyan et~al., {\it {Search for a charged Higgs
  boson decaying into a heavy neutral Higgs boson and a W boson in
  proton-proton collisions at $ \sqrt{s} $ = 13 TeV}},  {\em JHEP} {\bf 09}
  (2023) 032, [\href{http://arxiv.org/abs/2207.01046}{{\tt arXiv:2207.01046}}].

\bibitem{ATLAS:2024rcu}
{\bf ATLAS} Collaboration, G.~Aad et~al., {\it {Search for a heavy charged
  Higgs boson decaying into a W boson and a Higgs boson in final states with
  leptons and b-jets in $ \sqrt{s} $ = 13 TeV pp collisions with the ATLAS
  detector}},  {\em JHEP} {\bf 02} (2025) 143,
  [\href{http://arxiv.org/abs/2411.03969}{{\tt arXiv:2411.03969}}].

\bibitem{Li:2024kpd}
J.~Li, H.~Song, S.~Su, and W.~Su, {\it {Charged Higgs Search in 2HDM}},
  \href{http://arxiv.org/abs/2412.04572}{{\tt arXiv:2412.04572}}.

\bibitem{Branco:2011iw}
G.~C. Branco, P.~M. Ferreira, L.~Lavoura, M.~N. Rebelo, M.~Sher, and J.~P.
  Silva, {\it {Theory and phenomenology of two-Higgs-doublet models}},  {\em
  Phys. Rept.} {\bf 516} (2012) 1--102,
  [\href{http://arxiv.org/abs/1106.0034}{{\tt arXiv:1106.0034}}].

\bibitem{Paschos:1976ay}
E.~A. Paschos, {\it {Diagonal Neutral Currents}},  {\em Phys. Rev. D} {\bf 15}
  (1977) 1966.

\bibitem{Glashow:1976nt}
S.~L. Glashow and S.~Weinberg, {\it {Natural Conservation Laws for Neutral
  Currents}},  {\em Phys. Rev. D} {\bf 15} (1977) 1958.

\bibitem{Eriksson:2009ws}
D.~Eriksson, J.~Rathsman, and O.~Stal, {\it {2HDMC: Two-Higgs-Doublet Model
  Calculator Physics and Manual}},  {\em Comput. Phys. Commun.} {\bf 181}
  (2010) 189--205, [\href{http://arxiv.org/abs/0902.0851}{{\tt
  arXiv:0902.0851}}].

\bibitem{Bechtle:2020pkv}
P.~Bechtle, D.~Dercks, S.~Heinemeyer, T.~Klingl, T.~Stefaniak, G.~Weiglein, and
  J.~Wittbrodt, {\it {HiggsBounds-5: Testing Higgs Sectors in the LHC 13 TeV
  Era}},  {\em Eur. Phys. J. C} {\bf 80} (2020), no.~12 1211,
  [\href{http://arxiv.org/abs/2006.06007}{{\tt arXiv:2006.06007}}].

\bibitem{Bechtle:2020uwn}
P.~Bechtle, S.~Heinemeyer, T.~Klingl, T.~Stefaniak, G.~Weiglein, and
  J.~Wittbrodt, {\it {HiggsSignals-2: Probing new physics with precision Higgs
  measurements in the LHC 13 TeV era}},  {\em Eur. Phys. J. C} {\bf 81} (2021),
  no.~2 145, [\href{http://arxiv.org/abs/2012.09197}{{\tt arXiv:2012.09197}}].

\bibitem{Mahmoudi:2008tp}
F.~Mahmoudi, {\it {SuperIso v2.3: A Program for calculating flavor physics
  observables in Supersymmetry}},  {\em Comput. Phys. Commun.} {\bf 180} (2009)
  1579--1613, [\href{http://arxiv.org/abs/0808.3144}{{\tt arXiv:0808.3144}}].

\bibitem{Hahn:2001rv}
T.~Hahn and C.~Schappacher, {\it {The Implementation of the minimal
  supersymmetric standard model in FeynArts and FormCalc}},  {\em Comput. Phys.
  Commun.} {\bf 143} (2002) 54--68,
  [\href{http://arxiv.org/abs/hep-ph/0105349}{{\tt hep-ph/0105349}}].

\bibitem{Hahn:1998yk}
T.~Hahn and M.~Perez-Victoria, {\it {Automatized one loop calculations in
  four-dimensions and D-dimensions}},  {\em Comput. Phys. Commun.} {\bf 118}
  (1999) 153--165, [\href{http://arxiv.org/abs/hep-ph/9807565}{{\tt
  hep-ph/9807565}}].

\bibitem{Kublbeck:1990xc}
J.~Kublbeck, M.~Bohm, and A.~Denner, {\it {Feyn Arts: Computer Algebraic
  Generation of Feynman Graphs and Amplitudes}},  {\em Comput. Phys. Commun.}
  {\bf 60} (1990) 165--180.

\bibitem{Bahl:2022igd}
H.~Bahl, T.~Biek\"otter, S.~Heinemeyer, C.~Li, S.~Paasch, G.~Weiglein, and
  J.~Wittbrodt, {\it {HiggsTools: BSM scalar phenomenology with new versions of
  HiggsBounds and HiggsSignals}},  {\em Comput. Phys. Commun.} {\bf 291} (2023)
  108803, [\href{http://arxiv.org/abs/2210.09332}{{\tt arXiv:2210.09332}}].

\bibitem{Alwall:2014hca}
J.~Alwall, R.~Frederix, S.~Frixione, V.~Hirschi, F.~Maltoni, O.~Mattelaer,
  H.~S. Shao, T.~Stelzer, P.~Torrielli, and M.~Zaro, {\it {The automated
  computation of tree-level and next-to-leading order differential cross
  sections, and their matching to parton shower simulations}},  {\em JHEP} {\bf
  07} (2014) 079, [\href{http://arxiv.org/abs/1405.0301}{{\tt
  arXiv:1405.0301}}].

\bibitem{Hagiwara:2012vz}
K.~Hagiwara, T.~Li, K.~Mawatari, and J.~Nakamura, {\it {TauDecay: a library to
  simulate polarized tau decays via FeynRules and MadGraph5}},  {\em Eur. Phys.
  J. C} {\bf 73} (2013) 2489, [\href{http://arxiv.org/abs/1212.6247}{{\tt
  arXiv:1212.6247}}].

\bibitem{Sjostrand:2007gs}
T.~Sjostrand, S.~Mrenna, and P.~Z. Skands, {\it {A Brief Introduction to PYTHIA
  8.1}},  {\em Comput. Phys. Commun.} {\bf 178} (2008) 852--867,
  [\href{http://arxiv.org/abs/0710.3820}{{\tt arXiv:0710.3820}}].

\bibitem{Cacciari:2011ma}
M.~Cacciari, G.~P. Salam, and G.~Soyez, {\it {FastJet User Manual}},  {\em Eur.
  Phys. J. C} {\bf 72} (2012) 1896, [\href{http://arxiv.org/abs/1111.6097}{{\tt
  arXiv:1111.6097}}].

\bibitem{deFavereau:2013fsa}
{\bf DELPHES 3} Collaboration, J.~de~Favereau, C.~Delaere, P.~Demin,
  A.~Giammanco, V.~Lema\^\i{}tre, A.~Mertens, and M.~Selvaggi, {\it {DELPHES 3,
  A modular framework for fast simulation of a generic collider experiment}},
  {\em JHEP} {\bf 02} (2014) 057, [\href{http://arxiv.org/abs/1307.6346}{{\tt
  arXiv:1307.6346}}].

\bibitem{Cacciari:2008gp}
M.~Cacciari, G.~P. Salam, and G.~Soyez, {\it {The anti-$k_t$ jet clustering
  algorithm}},  {\em JHEP} {\bf 04} (2008) 063,
  [\href{http://arxiv.org/abs/0802.1189}{{\tt arXiv:0802.1189}}].

\bibitem{Conte:2013mea}
E.~Conte and B.~Fuks, {\it {MadAnalysis 5: status and new developments}},  {\em
  J. Phys. Conf. Ser.} {\bf 523} (2014) 012032,
  [\href{http://arxiv.org/abs/1309.7831}{{\tt arXiv:1309.7831}}].

\bibitem{Misiak:2020vlo}
M.~Misiak, A.~Rehman, and M.~Steinhauser, {\it {Towards $ \overline{B}\to
  {X}_s\gamma $ at the NNLO in QCD without interpolation in m$_{c}$}},  {\em
  JHEP} {\bf 06} (2020) 175, [\href{http://arxiv.org/abs/2002.01548}{{\tt
  arXiv:2002.01548}}].

\end{thebibliography}\endgroup
\end{document}